\documentclass{emulateapj}

\begin{document}

\shorttitle{The Initial-Final Mass Relation}
\shortauthors{Kalirai et al.}

\title{The Initial-Final Mass Relation: Direct Constraints at the 
Low Mass End\altaffilmark{1,2}}

\author{
Jasonjot S. Kalirai\altaffilmark{3,4}, 
Brad M.~S. Hansen\altaffilmark{5},
Daniel D. Kelson\altaffilmark{6},
David B. Reitzel\altaffilmark{5}, \\
R. Michael Rich\altaffilmark{5}, and
Harvey B. Richer\altaffilmark{7}
}
\altaffiltext{1}{Data presented herein were obtained at the W.\ M.\ Keck
Observatory, which is operated as a scientific partnership among the
California Institute of Technology, the University of California, and the
National Aeronautics and Space Administration.  The Observatory was made
possible by the generous financial support of the W.\ M.\ Keck Foundation.}
\altaffiltext{2}{Based on observations obtained at the Canada-France-Hawaii 
Telescope (CFHT) which is operated by the National Research Council of Canada, 
the Institut National des Sciences de l'Univers of the Centre National de 
la Recherche Scientifique of France, and the University of Hawaii.}
\altaffiltext{3}{University of California Observatories/Lick Observatory, 
University of California at Santa Cruz, Santa Cruz CA, 95060; jkalirai@ucolick.org}
\altaffiltext{4}{Hubble Fellow}
\altaffiltext{5}{Department of Physics and Astronomy, Box 951547, Knudsen Hall, 
University of California at Los Angeles, Los Angeles CA, 90095; 
hansen/rmr/reitzel@astro.ucla.edu}
\altaffiltext{6}{Carnegie Observatories, Carnegie Institution of Washington, 813 Santa 
Barbara Street, Pasadena CA, 91101; kelson@ociw.edu}
\altaffiltext{7}{Department of Physics and Astronomy, University of British Columbia, 
Vancouver, British Columbia, Canada, V6T 1Z1; richer@astro.ubc.ca}


\begin{abstract}

The initial-final mass relation represents a mapping between the mass of a 
white dwarf remnant and the mass that the hydrogen burning main-sequence 
star that created it once had.  The empirical relation thus far has been 
constrained using a sample of $\sim$40 stars in young open 
clusters, ranging in initial mass from $\sim$2.75 -- 7~$M_\odot$, and shows 
a general trend that connects higher mass main-sequence stars with higher 
mass white dwarfs.  In this paper, we present CFHT/CFH12K photometric and 
Keck/LRIS multiobject spectroscopic observations of a sample of 22 white 
dwarfs in two {\it older} open clusters, NGC 7789 ($t$ = 1.4~Gyr) and NGC 
6819 ($t$ = 2.5~Gyr).  At these ages, stars in these clusters with masses 
as low as 1.6~$M_\odot$ have already evolved off the main sequence and formed 
white dwarfs.  We measure masses for the highest S/N spectra by 
fitting the Balmer lines to atmosphere models and place the first direct constraints 
on the low mass end of the initial-final mass relation.  Our results indicate 
that the observed general trend at higher masses continues down to 
low masses, with $M_{\rm initial}$ = 1.6~$M_\odot$ main-sequence stars forming 
$M_{\rm final}$ = 0.54~$M_\odot$ white dwarfs.  When added to our new data 
from the very old cluster NGC~6791, the relation is extended down to 
$M_{\rm initial}$ = 1.16~$M_\odot$ (corresponding to $M_{\rm final}$ = 
0.53~$M_\odot$).  This extension of the relation represents a four fold 
increase in the total number of hydrogen burning stars for which the 
integrated mass loss can now be calculated from empirical data, assuming 
a Salpeter initial mass function.  The new leverage at the low mass end is 
used to derive a purely empirical initial-final mass relation for the entire sample 
of stars, without the need for any indirectly measured anchor points.  The sample of 
white dwarfs in these clusters also shows several very interesting systems that 
we discuss further: a DB (helium atmosphere) white dwarf, a magnetic white dwarf, 
a DAB (mixed hydrogen/helium atmosphere or a double degenerate DA+DB) white 
dwarf(s), and two possible equal mass DA double degenerate binary systems.

\end{abstract}



\section{Introduction} \label{introduction}

The initial-final mass relation denotes a mapping from the initial 
mass of a main-sequence star to its final white dwarf configuration and 
hence provides the total mass loss that a star has undergone through its 
lifetime, a fundamental property of stellar evolution \citep{reimers75,renzini88,weidemann00}.  
At one extreme, a small extrapolation of the high mass end of the relation 
can lead to constraints on the critical mass that separates white dwarf 
production from type II supernova explosions.  This can therefore be used 
to estimate energetics involved in feedback processes through the prediction 
of the birth rates of type II supernovae and neutron stars \citep{vandenbergh91}.  
At the opposite extreme, the relation represents a rare tool to probe the 
progenitor properties of the majority of the evolved stars in old stellar 
populations (most of which are now low mass white dwarfs).  If constrained 
over a large mass range (i.e., 1 -- 7 $M_\odot$), the relation can be a 
powerful input to chemical evolution models of galaxies (including enrichment 
in the interstellar medium) and therefore enhances our understanding of 
star formation efficiencies in these systems \citep{somerville99}.

The importance of the initial-final mass relation has been recently 
compounded as a result of the discovery of thousands of white dwarfs 
in both the Galactic disk and halo.  For the former, the Sloan Digital 
Sky Survey has spectroscopically confirmed many new white dwarfs 
bringing the total number of such objects in our Galaxy to almost 10,000 
\citep{eisenstein06}.  This has led to an improved white luminosity 
function for the disk of our Galaxy that shows an abrupt truncation 
at $M_{\rm bol}$ = 15.3 \citep{harris06}.  In the Galactic halo, recent 
{\it Hubble Space Telescope} observations of the globular clusters 
M4 \citep{richer04,hansen04}, Omega~Cen \citep{monelli05}, and NGC~6397 
\citep{richer06,hansen07} have similarly uncovered several thousand 
cluster white dwarfs.  Modeling the luminosity functions of the disk 
white dwarfs and the cooling sequences of the halo star clusters, directly 
yields the ages of the Galactic disk and halo components.  In both cases, 
the white dwarf samples are dominated by low mass stars and therefore 
such modeling requires an input initial-final mass relation that is 
well understood at the low mass end (e.g., Ferrario et~al.\ 2005 and 
Hansen et~al.\ 2007).

The first attempt to derive an initial-final mass relation was made by 
\cite{weidemann77}.  He compared theoretical models of mass loss 
(e.g., Fusi-Pecci \& Renzini 1976) to the observed masses of a few 
white dwarfs in the nearby Hyades and Pleiades clusters and 
concluded that the observed mass loss was larger than model predictions.  
Shortly after this pioneering work, \cite{romanishin80} 
and \cite{anthonytwarog81,anthonytwarog82} used photographic plates to 
search for new white dwarf candidates in several young open clusters, 
including NGC~1039, NGC~2168, NGC~2287, NGC~2422, NGC~2632 (Praesepe), 
NGC~6633, NGC~6405, and IC~2602.  These studies modeled the expected 
numbers of white dwarfs in each cluster and estimated limits on the 
boundaries for the upper progenitor mass limit to white dwarf production 
(5 -- 7~$M_\odot$).  Solid constraints on the relation came from subsequent 
spectroscopic observations of these white dwarfs as well as newly 
discovered degenerate stars in nearby open clusters (Koester \& Reimers 1981, 
1985, 1993, 1996; Reimers \& Koester 1982, 1989, 1994; Weidemann \& 
Koester 1983; Weidemann 1987, 1997; Jeffries 1997).  The result of this enormous 
two-decade long effort was an initial-final mass relation consisting of 
$\sim$20 data points, from observations of roughly a half-dozen open 
star clusters (see Weidemann~2000 for a review).  The final relation 
shows a clear trend with higher mass main-sequence stars producing 
increasingly more massive white dwarfs. 

In the last few years, the amount of data constraining the initial-final 
mass relation has more than doubled 
\citep{claver01,dobbie04,dobbie06,williams04,kalirai05a,liebert05b,williams07}.  
Although the general trend of the relation remains intact, the scatter 
has increased possibly signifying a relation between the stellar mass 
loss and the properties of the host environment (e.g., metallicity effects 
-- Kalirai et~al.\ 2005a).  As an extreme example, the recent study of 
the white dwarf population of the super-solar metallicity star cluster 
NGC~6791 ([Fe/H] = $+$0.4) has revealed it to be significantly undermassive 
relative to the field distribution.  This is clear evidence that the progenitor 
stars of these remnants experienced enhanced mass loss in post main-sequence 
evolutionary stages due to the high metallicity of the cluster \citep{kalirai07}.  

Prior to this study, the oldest open star clusters that have been successfully targeted 
for white dwarf spectroscopy to build an initial-final mass relation are the Hyades and 
Praesepe\footnote{Fleming et~al.\ (1997) also discuss one object along M67's 
sightline whose membership remains uncertain.}.  The ages of both of these systems are 
600 -- 700~Myr \citep{perryman98,claver01}, 
indicating that the present day turn-off masses are $\gtrsim$2.75~$M_\odot$.  This 
threshold therefore represents the current low mass anchor on the initial-final 
mass relation as all of the white dwarfs in these clusters must have evolved from 
main-sequence progenitors with a mass larger than $\sim$2.75~$M_\odot$.  A very 
small fraction of all stars in the Universe have masses this large, and therefore 
the relation is often extrapolated to lower masses to provide useful input.  
Spectroscopic white dwarf studies have been unable to target any old open clusters 
($t >$ 1~Gyr) for several reasons.  Primarily, few photometric studies exist that 
have identified populations of white dwarf candidates in these clusters.  Second, 
the known rich old open clusters are generally much further ($>$10$\times$) 
than nearby clusters such as the Hyades and Praesepe.  Finally, because of 
their older age, most cluster white dwarfs in these systems have cooled to very 
faint magnitudes thus making it difficult to obtain high quality spectra of the 
stars.  

White dwarfs in the nearest globular star clusters have also recently 
been targeted for mass measurements by several groups.  The only successful 
campaign measured a mean mass of 0.53~$M_\odot$ for white dwarfs in NGC~6752 
\citep{moehler04}.  Given the lower S/N of these data, the 
temperature of the stars was measured from the spectra and then combined 
with photometric information to yield a mass.  This mean mass is consistent 
with several independent arguments that all suggest the masses 
of white dwarfs in globular clusters should be 0.51 -- 0.55 $M_\odot$ 
\citep{renzini88,renzini96}.


\begin{table}
\begin{center}
\caption{}
\begin{tabular}{lccr}
\hline
\hline
\multicolumn{1}{c}{Filter} & \multicolumn{1}{c}{Exp. Time (s)} & 
\multicolumn{1}{c}{No. Images}  & \multicolumn{1}{c}{Airmass} \\ 
\hline

{\bf NGC~7789} \\

$V$ & 600  & 14 & 1.25 -- 1.43 \\
$V$ & 90   & 1  & 1.67 \\ 
$V$ & 10   & 1  & 1.25 \\
$V$ & 5    & 1  & 1.28 \\
$V$ & 1    & 1  & 1.73 \\
$V$ & 0.5  & 1  & 1.69 \\
$B$ & 800 & 12  & 1.25 -- 1.42 \\
$B$ & 120 & 1 & 1.35 \\
$B$ & 10  & 1 & 1.36 \\
$B$ & 1   & 1 & 1.37 \\
$B$ & 0.5 & 1 & 1.37 \\

{\bf NGC~6819} \\

$V$ & 300 & 9 & 1.16 -- 1.31 \\
$V$ & 50  & 1 & 1.16 \\
$V$ & 10  & 1 & 1.15 \\
$V$ & 1   & 1 & 1.27  \\
$B$ & 300 & 9 & 1.40 -- 1.76 \\
$B$ & 50  & 1 & 1.38 \\
$B$ & 10  & 1 & 1.37 \\
$B$ & 1   & 1 & 1.25   \\
\hline
\end{tabular}
\label{table1}
\end{center}
\end{table}



\begin{figure*}
\begin{center}
\leavevmode 
\includegraphics[height=16.5cm,angle=270]{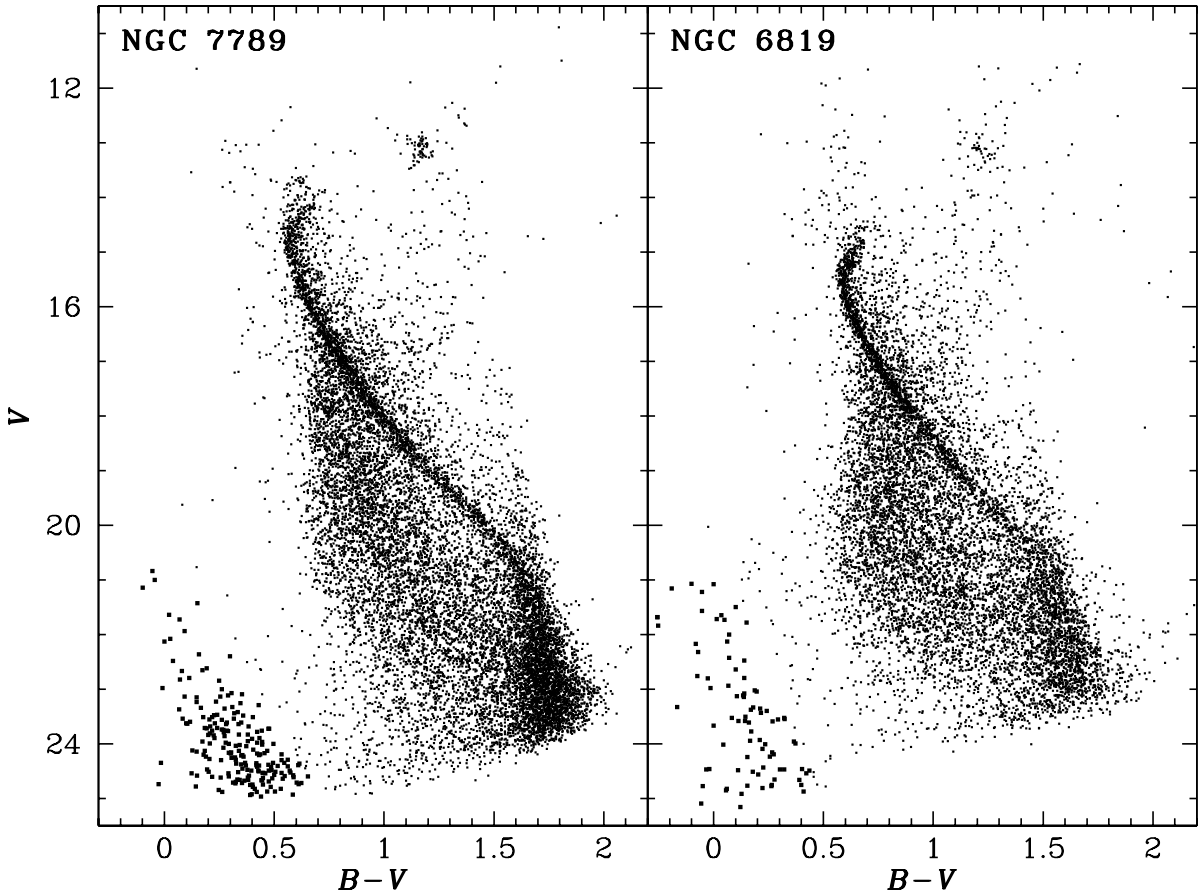}
\end{center}
\caption{The CMDs of NGC~7789 and NGC~6819 show very tight 
main-sequences, turnoffs, and post-main sequence evolutionary 
phases. For example, a ``hook''  is seen above the 
turnoff designating the contraction of stars that have just 
exhausted their hydrogen supply.  These are the deepest CMDs 
constructed for these clusters to date and the faint-blue 
region of the CMDs reveals a large population of white dwarfs 
in each cluster (see also Kalirai et~al.\ 2001b).
\label{fig:2cmdsnoiso}}
\end{figure*}


The combination of large mosaic cameras on 4-meter telescopes (e.g., 
CFH12K/MegaCam on the Canada-France-Hawaii Telescope) and the advent of 
blue-sensitive multiobject spectrographs on 10-meter telescopes (e.g., LRIS 
on Keck -- Oke et~al.\ 1995) provide the resources necessary to extend 
the study of the initial-final mass relation to a new regime.  
In this paper we present direct spectroscopic mass determinations of white dwarfs 
in open clusters older than 1~Gyr.  The very rich clusters NGC~7789 and 
NGC~6819 are $\sim$2$\times$ and $\sim$4$\times$ older than the 
Hyades/Praesepe systems, respectively, and have been recently studied 
by our team using the Canada-France-Hawaii Telescope to very 
faint magnitudes ($V \sim$ 25).  The data have uncovered hundreds of 
white dwarf candidates which have been followed up with the Keck 10-meter 
telescope and LRIS multiobject spectrograph.  In the following 
section we present our photometric observations of NGC~7789 and 
NGC~6819 and construct the deepest color-magnitude diagrams (CMDs) for 
each cluster to date (\S\,3).  Parameters (e.g., distance, reddening, and 
age) are derived for each cluster.  In \S\,4 we discuss the construction of 
multi-object spectroscopic masks to observe the candidate white dwarfs 
in each cluster and describe the general spectroscopic observations.  This 
includes the selection of white dwarf candidates from the imaging catalogs.  
The spectra for all confirmed DA (hydrogen atmosphere) white dwarfs 
are presented in \S\,5 and fit to synthetic spectra to derive 
$T_{\rm eff}$, log~$g$, masses, and cooling ages in \S\,6.  We eliminate 
field white dwarfs from our sample and calculate the progenitor masses 
for each of the cluster white dwarfs in \S\,7.  This is used to 
build a new empirical initial-final mass relation extending down 
to $M_{\rm initial}$ = 1.6~$M_\odot$.  When added to our recent study of 
the 8.5~Gyr cluster NGC~6791, the relation is now mapped down to 
$M_{\rm initial}$ = 1.16~$M_\odot$.  The results are presented and 
analyzed in \S\,8 and the study is summarized in \S\,9.

\section{CFHT Photometry} \label{imagingobservations}

All of the imaging observations of NGC~7789 and NGC~6819 
were obtained with the CFH12K mosaic CCD camera on the 
4-meter Canada-France-Hawaii Telescope (CFHT), as a part of 
the CFHT Open Star Cluster Survey \citep{kalirai01a}.  The 
camera contains 12 CCDs, each with 2048 $\times$ 4096 
pixels (a total of over 100 million pixels), at an individual 
pixel scale of 0$\farcs$206.  The projection on the sky is 
42$'$ $\times$ 28$'$ and therefore the dominant population of 
both clusters is probed out to near the tidal radii.

We imaged NGC~7789 from late May to mid July 2001 in the $V$ and 
$B$ filters.  Similarly, NGC~6819 data were acquired in the 
same filters in October 1999, April 2001, and August 2001.  
Multiple deep exposures were taken to achieve a solid detection 
of the white dwarf cooling sequence in each cluster (no previous 
white dwarfs had been found in either system).  Shallower 
exposures were also obtained to fill in the brighter 
main-sequence, turnoff, and giant stars, which are saturated on the 
longer frames.  In all exposures, the clusters were placed 
near the center of the mosaic camera to allow a suitable blank 
field to be constructed from the outer CCDs.  

The observing conditions were very good for the majority of the 
exposures (photometric skies, sub-arcsecond seeing 
conditions, and low airmasses).  A log of the data used 
in the final analysis is presented in Table~1.  

The data reduction for NGC~6819 is described in detail in 
\cite{kalirai01b}.  The final photometric and astrometric 
catalogs used in the present study are identical to that earlier set.  
For NGC~7789, we processed the science frames according to the 
prescription in \cite{kalirai01a}.  Summarizing, we obtained several 
flat-field, bias, and dark images and applied these to the 
individual science frames using the FITS Large Images Processing 
Software\footnote{\url{http://www.cfht.hawaii.edu/$\sim$jcc/Flips/flips.html}} 
(FLIPS -- see also Kalirai et~al.\ 2001a).  FLIPS was 
next used to register and coadd the multiple science exposures 
for a given exposure time (which were each dithered slightly).  
Photometry was performed on the resulting images using a 
variable point-spread function in DAOPHOT \citep{stetson94}, 
and calibrated using Landolt standard star field observations \citep{landolt92} 
as discussed in \S\S~5.1 and 5.2 of \cite{kalirai01a}.  The 
final errors in the photometry are very well behaved. We 
find $\sigma_V$ $<$ 0.05 mag down to $V$ = 24 and 
$\sigma_V$ $<$ 0.10 mag down to $V$ = 25.


\section{Color-Magnitude Diagrams} \label{CMD}

NGC~7789 and NGC~6819 are two of the richest open star clusters 
in the Milky Way.  As we show below, the clusters are both old and 
located at a similar distance from the Sun.  The positions of these 
systems in the Galaxy is also quite similar; they are both 
found within 10 degrees of the plane of the Galactic disk at 
$l$ = 115.5$^{\rm \circ}$ (NGC~7789) and $l$ = 74.0$^{\rm \circ}$ (NGC~6819).  
Not surprisingly, the CMDs of the two clusters are strikingly similar 
as shown in Figure~\ref{fig:2cmdsnoiso}.  The contrast of the 
cluster main sequences over the foreground and background Milky Way 
disk and halo populations is very strong.  In the observational plane, the main 
sequence can be seen extending from the bluest point on the present day 
turnoff ($V \sim$ 14.6 and $B-V$ $\sim$ 0.6 for NGC~7789, $V \sim$ 15.4 and 
$B-V$ $\sim$ 0.6 for NGC~6819) down to the photometric limit.  Just to the 
red of the main sequence, an equal mass binary sequence can be seen in 
NGC~6819 and also possibly in NGC~7789.  As we showed through synthetic 
CMD fitting in \cite{kalirai04}, the fraction of binary stars in NGC~6819 
is $\sim$20 -- 30\%.  The turnoff of both clusters is clearly defined 
as well as an apparent ``hook'' just above the brightest point.  This hook 
is caused by a small contraction of the stellar core just above the 
turnoff.  Few, if any, stars are seen in the subgiant branch given the 
shorter evolutionary timescale of this phase of post main-sequence 
evolution.  However, several blue-straggler candidates are found above 
the turnoff in both clusters, especially in NGC~6819.  The horizontal 
branches are manifested as red clumps, as expected given the higher metallicity 
of stars in these two systems.  Evolution off the red giant clump is also seen 
at the bright-red part of the CMDs.


\begin{figure}
\epsscale{1.1} \plotone{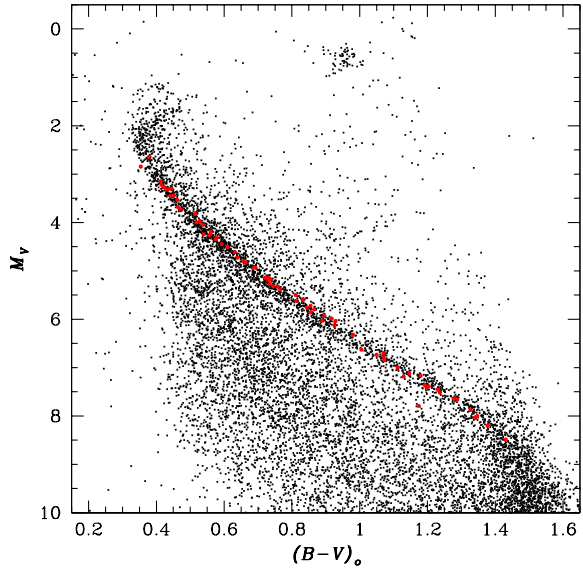} 
\figcaption{The distance modulus of NGC~7789 is measured to be 
($m-M$)$_V$ = 12.5 $\pm$ 0.1 by matching the observed cluster main 
sequence (blue edge) to the Hyades cluster (red dots).  The slope of 
the two main-sequences are in excellent agreement over the entire 
CMD.  In this plane, all adjustments (distance, reddening, and color 
offset due to metallicity difference) have been made to the NGC~7789 
stars (see \S\,\ref{red.dist.age}).
\label{fig:hyadesfig}}
\end{figure}


These CMDs of NGC~7789 and NGC~6819 are the deepest ever constructed for 
the clusters (note: NGC~6819 was presented in Kalirai et~al.\ 2001b).  
As expected, each CMD shows a large population of white dwarfs that had 
previously not been detected.  These 
cooling sequences, in the faint-blue part of the diagrams, extend several 
magnitudes to the limit of the data (these points have been made larger for 
clarity).  Unlike our study of the 0.5 Gyr cluster NGC~2099 \citep{kalirai01c}, 
these two clusters are old enough that the coolest white dwarfs 
($V$ = 26 -- 27) are beyond our detection limit.  The scatter in the 
cooling sequences results from a combination of photometric errors and 
field contamination, which we will address later in section~\ref{membership}.

\subsection{Distance and Age Measurements} \label{red.dist.age}

With just two color photometry, it is very difficult to simultaneously 
constrain the reddening and distance of a star cluster.  When fitting the 
main sequence, these parameters are degenerate.  Fortunately, the 
reddening can be measured independently from multi-filter photometry.  
\cite{wu07} recently presented a 13 color CCD spectrophotometric study of 
NGC~7789 and conclude with an estimate of the foreground reddening to NGC~7789 of 
E($B-V$) = 0.28 $\pm$ 0.02.  In their Table 1, they also list previous 
measurements (dating back to the work of Burbidge \& Sandage 1958) 
and find that their value is in fact nicely bracketed by 
the findings in these independent studies (0.22~$<$~E($B-V$)~$<$~0.35, 
see references within Wu et~al.\ 2007).  

The cornerstone technique of determining the distance of an open star 
cluster involves fitting the observed main sequence to that of the 
Hyades cluster.  As the nearest star cluster to the Sun ($d$ = 46.34 $\pm$ 0.27~pc 
-- Perryman et~al.\ 1998), the distance to each of the Hyades main-sequence 
stars is accurately known through parallax measurements (to within 
$\sim$2\% -- de~Bruijne, Hoogerwerf, \& de~Zeeuw 2001).  Therefore, one 
can directly overlay the Hyades main sequence stars (in an $M_V$, $(B-V){\rm o}$ 
plane) to the observed cluster main-sequence and adjust the distance 
modulus of the latter until the two overlap.  Although NGC~7789 is much 
older than the Hyades (more than a factor of two), our deep photometry 
presents a long, unevolved main sequence for this comparison.  We do 
however need to make a slight adjustment given the different metallicities 
of the clusters.  The Hyades is slightly more metal-rich than the Sun, 
$Z$ = 0.024 \citep{perryman98}, whereas NGC~7789 is slightly more metal-poor, 
$Z$ = 0.014 (average of recent literature values, see Wu et~al.\ 2007).  
Correcting this offset amounts to a very small color shift of the 
main-sequence.  The resulting comparison yields an excellent alignment 
of the two main sequences for an NGC~7789 distance modulus of ($m-M$)$_V$ 
= 12.5 $\pm$ 0.1, where the error bar is derived as described in section 
8.4 of \cite{kalirai01c}.  This is illustrated in Figure~\ref{fig:hyadesfig} 
where we have overlaid the Hyades stars on top of the shifted NGC~7789 main 
sequence. 

Using a similar analysis, \cite{kalirai01b} determined the reddening and 
distance modulus of NGC~6819 to be E($B-V$) $\sim$ 0.10 -- 0.14 (see also 
Bragaglia et~al.\ 2001) and ($m-M$)$_V$ = 12.30 $\pm$ 0.12.





\begin{figure}
\epsscale{1.1} \plotone{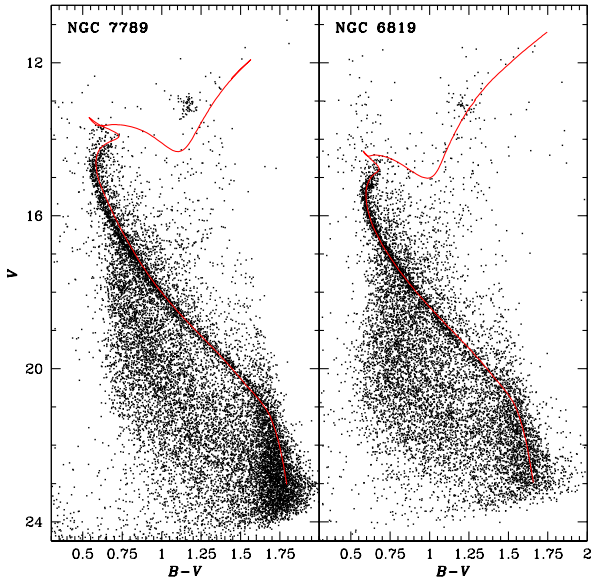} 
\figcaption{Stellar isochrones from \cite{vandenberg06} are found to be in 
excellent agreement with the main sequence and main-sequence 
turnoff of NGC~7789 and NGC~6819 (although there is a disagreement on the 
red giant branch of NGC~7789).  As summarized in section \ref{red.dist.age}, 
we measure an age of $t$ = 1.4~Gyr for NGC~7789 and $t$ = 2.5~Gyr for NGC~6819.
\label{fig:2cmdsiso}}
\end{figure}


With an estimate of the fundamental parameters in place, we can measure 
the ages of both clusters using our derived CMDs.  For this, we have chosen 
to use the stellar isochrones from \cite{vandenberg06} which include a 
more physical treatment of convective overshooting than past generation 
models (see below for a comparison with other models).  Our results are 
shown in Figure~\ref{fig:2cmdsiso}.  Assuming 
[$\alpha$/Fe] = 0, our CMD for NGC~7789 favors an isochrone with $t$ = 
1.4~Gyr.  The resulting fit to the entire main-sequence and turnoff 
is good, although the cluster red giants are bluer than the model 
prediction.  We note that \cite{vandenberg06} also found this discrepancy 
when fitting an older photometric data set ($V$, $I$) of this cluster (observed 
by Gim et~al.\ 1998).  The cause of this mismatch may be in part related to 
the masses of these red giant stars, which, given the age of NGC~7789, should 
be very close to the phase transition threshold where the evolution is 
terminated by degenerate helium ignition in the core (i.e., the flash).  
Modeling this transition depends sensitively on the extent of core 
overshooting.  For NGC~6819, we find that an isochrone of age 
$t$ = 2.5~Gyr reproduces all of the main CMD features very nicely.  This 
includes the main-sequence, turnoff, and red giant branch.  For both 
clusters, the ages determined from the \cite{vandenberg06} isochrones 
are consistent at the $\sim$10\% level with those determined from either 
the Yale-Yonsei isochrones \citep{demarque04} or the Padova group 
isochrones \citep{girardi02}.


\begin{figure}
\epsscale{1.1} \plotone{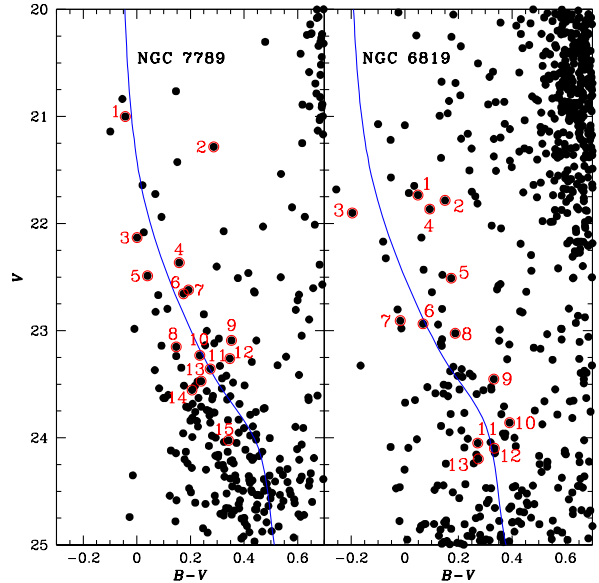} 
\figcaption{The 28 white dwarf candidates that are spectroscopically targeted 
in NGC~7789 and NGC~6819 are highlighted in the faint-blue corner of the 
cluster CMDs.  These objects are scattered around a 0.6~$M_\odot$ cooling 
sequence \citep{wood95} and span approximately three magnitudes of the 
white dwarf cooling sequence in each cluster. 
\label{fig:wdzoom}}
\end{figure}


To summarize the analysis of the cluster CMDs, our best parameters 
are E($B-V$) = 0.28 $\pm$ 0.02, ($m-M$)$_V$ = 12.5 $\pm$ 0.1, 
$Z$ = 0.014, and $t$ = 1.4~Gyr for NGC~7789.  For NGC~6819, we 
find E($B-V$) = 0.13 $\pm$ 0.02, ($m-M$)$_V$ = 12.30 $\pm$ 0.12, $Z$ = 0.017, 
and $t$ = 2.5~Gyr.  We point out that these age derivations are 
sensitive to the input parameters.  A reasonable fit to the observed 
CMDs can be achieved by tweaking the age by $\sim$10\% with corresponding 
changes to the reddening, distance moduli, and/or metallicity.  Although 
we can not be absolutely certain which combination of these parameters are 
correct for the clusters (given the ranges reported in the literature), 
we are reasonably sure that our estimates are accurate as they agree 
with most recent literature values.  It is also reassuring that, for 
this set of parameters, the models reproduce the lower main sequences 
nearly perfectly (this phase has never been tested before).


\begin{figure*}
\begin{center}
\leavevmode 
\includegraphics[height=18.5cm,angle=90]{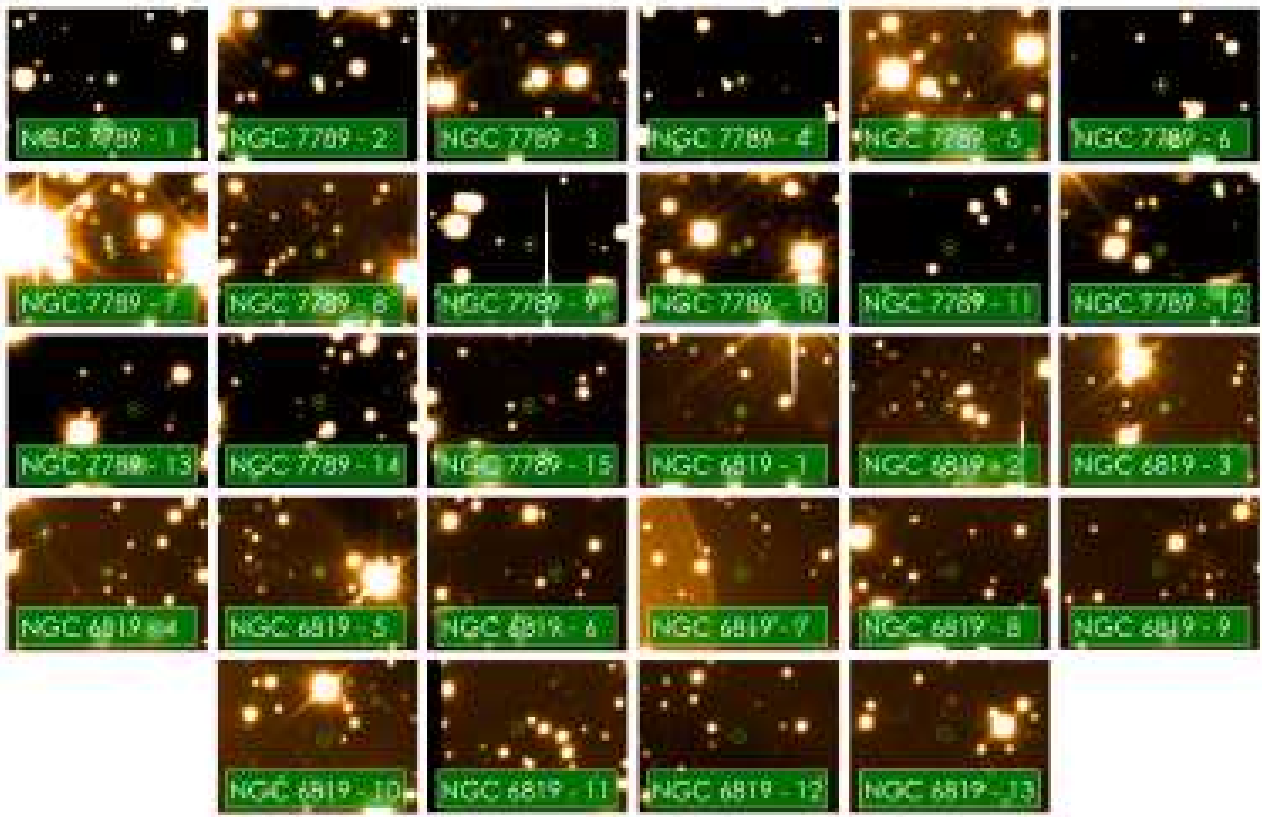}
\end{center}
\caption{The images of each of our white dwarf candidates are 
shown from the $V$-band CFHT data.  Each star is displayed in 
a small window that extends approximately 1 arcminute in the 
E-W direction (E is to the left) and 35 arcseconds in the 
N-S direction (N is to the top).  By design, most of the white dwarf 
candidates are well isolated, sharp sources.  Higher resolution 
version of this figure is available at 
http://www.ucolick.org/$\sim$jkalirai/0706.3894/. \label{fig:snapshot}}
\end{figure*}




\section{Keck Spectroscopy} \label{spectroscopicdata}

Spectroscopic observations of NGC~7789 and NGC~6819 were obtained using 
the LRIS multi-object spectrograph on the Keck~I telescope on
July~29 and July~30 2005.  The instrument is a dual-beam, low resolution 
spectrograph with a 5$' \times$ 7$'$ field of view \citep{oke95}.  For 
the blue side, we used the 600/4000 grism (dispersion = 0.63 ${\rm \AA}$/pixel) 
which simultaneously covers 2580~${\rm \AA}$, from 3300 -- 5880~${\rm \AA}$.  The plate 
scale of the blue CCD is 0$\farcs$135 per pixel.  For the 
red side, we used the 600/7500 grating (dispersion = 1.28 ${\rm \AA}$/pixel), 
centered at 6600~${\rm \AA}$, which covers a wavelength baseline of 
2620~${\rm \AA}$.  The plate scale of the red CCD is 0$\farcs$210 per 
pixel.  The light to the blue side was intercepted from the 
collimator mirror using the D560 dichroic.  In multiobject 
slit spectroscopy, the exact wavelength coverage for each target varies 
somewhat depending on the location of that target on the mask. 

We do not a priori know {\it which} of the faint-blue stars identified as 
white dwarf candidates from the imaging observations are in fact white dwarfs.  
Unresolved background galaxies, QSOs, hot subdwarfs, and even distant early 
type main-sequence stars can contaminate the sample.  However, NGC~7789 and 
NGC~6819 are two of 
the richest Milky Way open star clusters and therefore the percentage of 
contaminating field objects is suppressed.  In fact, Figure~\ref{fig:2cmdsnoiso} 
shows that both clusters exhibit obvious white dwarf cooling sequences which 
would not be otherwise discernible if field contamination was overwhelming.  We 
also note that similar studies by our group of the rich cluster NGC~2099 
\citep{kalirai05a} and NGC~6791 \citep{kalirai07} have confirmed that most 
faint-blue objects in our CFHT CMDs for these rich systems are in fact 
cluster white dwarf members.


\begin{table*}
\begin{center}
\caption{}
\begin{tabular}{lcccr}
\hline
\hline
\multicolumn{1}{c}{ID} & \multicolumn{1}{c}{$\alpha_{J2000}$} & \multicolumn{1}{c}{$\delta_{J2000}$} & 
\multicolumn{1}{c}{$V$} & \multicolumn{1}{c}{$B-V$} \\
\hline
NGC 7789~--~1  & 23:56:32.51 & 56:35:21.6 & 21.00 $\pm$ 0.01  & $-$0.04 \\
NGC 7789~--~2  & 23:56:44.25 & 56:38:09.7 & 21.28 $\pm$ 0.01  &    0.29 \\
NGC 7789~--~3  & 23:56:36.62 & 56:40:23.5 & 22.13 $\pm$ 0.02  &    0.01 \\
NGC 7789~--~4  & 23:56:43.03 & 56:35:56.2 & 22.37 $\pm$ 0.02  &    0.16 \\
NGC 7789~--~5  & 23:56:49.06 & 56:40:13.2 & 22.49 $\pm$ 0.01  &    0.04 \\
NGC 7789~--~6  & 23:56:31.94 & 56:36:59.2 & 22.66 $\pm$ 0.02  &    0.17 \\
NGC 7789~--~7  & 23:56:51.93 & 56:38:21.3 & 22.62 $\pm$ 0.01  &    0.19 \\
NGC 7789~--~8  & 23:56:57.22 & 56:40:01.1 & 23.15 $\pm$ 0.02  &    0.15 \\
NGC 7789~--~9  & 23:56:42.42 & 56:32:48.4 & 23.09 $\pm$ 0.02  &    0.35 \\
NGC 7789~--~10 & 23:56:44.91 & 56:39:58.8 & 23.23 $\pm$ 0.02  &    0.24 \\
NGC 7789~--~11 & 23:56:30.81 & 56:37:19.3 & 23.36 $\pm$ 0.02  &    0.27 \\
NGC 7789~--~12 & 23:56:45.84 & 56:37:55.1 & 23.26 $\pm$ 0.02  &    0.35 \\
NGC 7789~--~13 & 23:57:05.17 & 56:38:20.1 & 23.47 $\pm$ 0.02  &    0.24 \\
NGC 7789~--~14 & 23:56:37.78 & 56:39:08.4 & 23.55 $\pm$ 0.02  &    0.21 \\
NGC 7789~--~15 & 23:56:34.19 & 56:40:05.0 & 24.02 $\pm$ 0.03  &    0.34 \\
NGC 6819~--~1  & 19:41:25.70 & 40:02:53.9 & 21.73 $\pm$ 0.01  &    0.05 \\  
NGC 6819~--~2  & 19:41:26.10 & 40:03:48.0 & 21.78 $\pm$ 0.01  &    0.15 \\ 
NGC 6819~--~3  & 19:41:48.06 & 40:03:17.0 & 21.90 $\pm$ 0.01  & $-$0.20 \\
NGC 6819~--~4  & 19:41:46.80 & 40:03:08.2 & 21.87 $\pm$ 0.01  &    0.09 \\
NGC 6819~--~5  & 19:41:27.36 & 40:00:47.8 & 22.51 $\pm$ 0.01  &    0.17 \\
NGC 6819~--~6  & 19:41:19.96 & 40:02:56.1 & 22.94 $\pm$ 0.02  &    0.07 \\
NGC 6819~--~7  & 19:41:33.93 & 40:01:41.4 & 22.91 $\pm$ 0.02  & $-$0.02 \\
NGC 6819~--~8  & 19:41:37.21 & 40:04:45.3 & 23.03 $\pm$ 0.02  &    0.19 \\
NGC 6819~--~9  & 19:41:25.20 & 40:01:30.2 & 23.45 $\pm$ 0.03  &    0.33 \\
NGC 6819~--~10 & 19:41:32.13 & 40:01:07.8 & 23.86 $\pm$ 0.04  &    0.39 \\
NGC 6819~--~11 & 19:41:53.95 & 40:04:05.7 & 24.05 $\pm$ 0.05  &    0.27 \\
NGC 6819~--~12 & 19:41:32.76 & 40:04:39.9 & 24.10 $\pm$ 0.06  &    0.33 \\
NGC 6819~--~13 & 19:41:32.18 & 40:05:59.7 & 24.20 $\pm$ 0.06  &    0.27 \\
\hline
\end{tabular}
\label{table2}
\end{center}
\end{table*}


We generate an input list of spectroscopic targets by assigning priorities 
to objects in the CFHT CMD based on their magnitudes and morphology (i.e., 
extended sources with a poor ``stellarity'' are removed -- Bertin \& Arnouts 
1996).  Objects that are near the bright white dwarf cooling sequence (defined 
by eye, see Figure~\ref{fig:2cmdsnoiso}) in each cluster are given high priorities 
and objects that are fainter are given lower priorities.  Since the LRIS 
field of view is much smaller than our wide-field CFHT image, we strategically 
position the 
spectroscopic mask to overlap as many of the best targets as we can.  Our 
expectation was to observe a single field in each of the clusters to maximize 
the S/N of the resulting spectra, which is critical 
to derive accurate masses (see \S\,\ref{WDMasses}).  However, we generated 
spectroscopic masks at two different locations in case a quick reduction of 
the data from the first exposure taken at the telescope revealed that most 
of the targets were not white dwarfs.  In this case, we had the option to 
abandon further exposures of that particular field and switch to the second 
mask which targeted a different region of the cluster.


\begin{figure*}
\begin{center}
\leavevmode 
\includegraphics[height=17.0cm,angle=270]{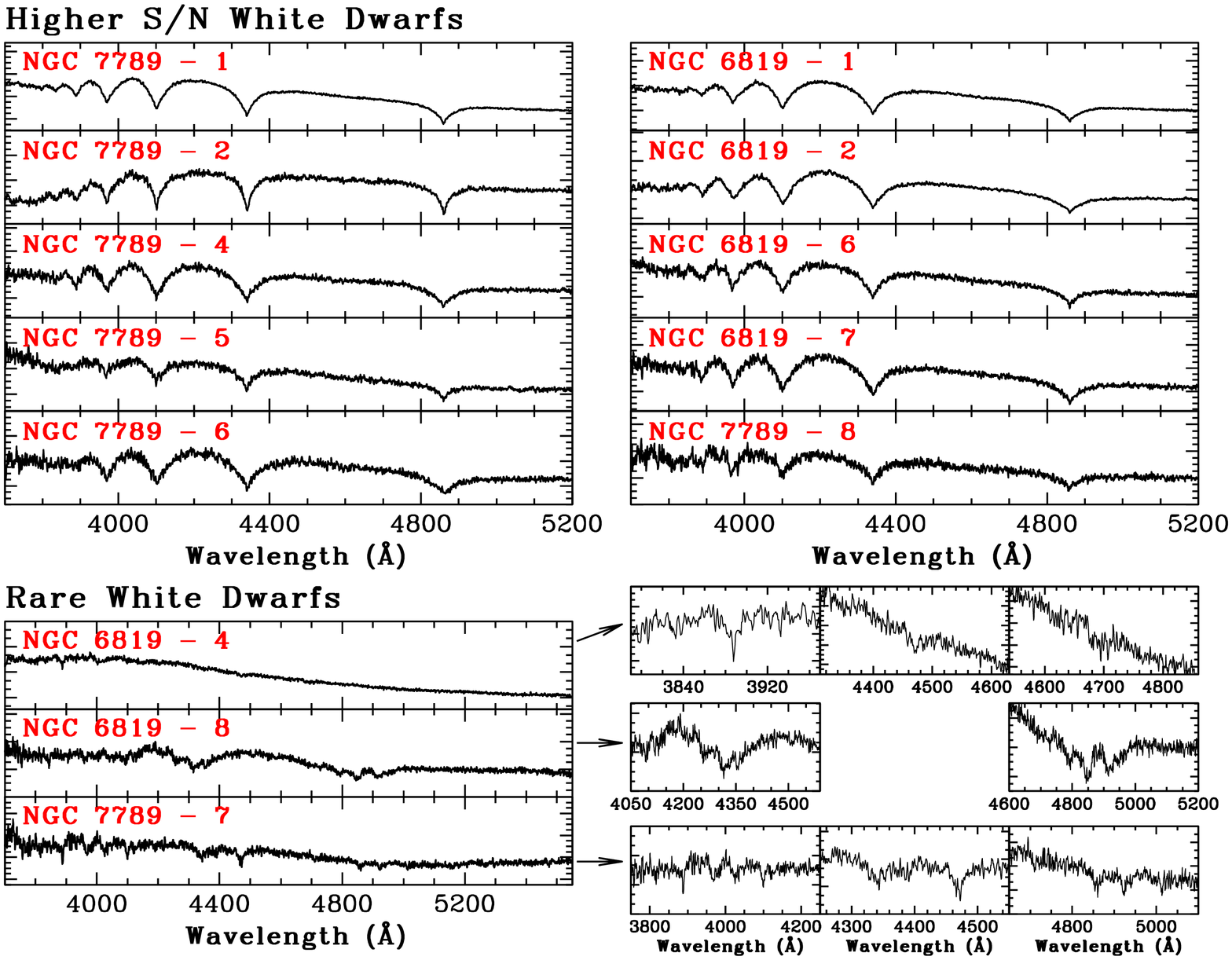}
\end{center}
\caption{{\it Top} -- Keck/LRIS spectra for ten stars with at least 
five well characterized Balmer lines.  Each of these stars can be fit 
to models to yield accurate temperatures and gravities (see section~\ref{WDMasses}). 
{\it Bottom} -- The spectra of three unique white dwarfs ({\it left}) 
and a closer look at their absorption lines (NGC~6819~--~4 and 8, 
and NGC~7789~--~7 -- {\it right}).  NGC~6819~--~4 shows the presence of 
He lines at 3889, 4471, and 4713 ${\rm \AA}$ and is therefore a DB (helium 
atmosphere) white dwarf.  NGC~6819~--~8 shows obvious signatures of 
very broad H$\beta$ and H$\gamma$ Balmer lines with possible Zeeman 
splitting, but no clear evidence of higher order Balmer lines.  
This star looks to be a massive, magnetic white dwarf.  The spectrum of 
NGC~7789~--~7 shows both hydrogen and helium absorption lines.   These 
objects are discussed further in section~\ref{rareobjects}.
\label{fig:wdspectrahigh}}
\end{figure*}


For each mask location discussed above (and similarly two locations in 
NGC~6819), we milled two masks with individual slit widths of 
0$\farcs$8 and 1$\farcs$0, and orientations close to the parallactic angle.  
The choice between the two masks was made dependent on the seeing conditions 
of the observations.  The individual exposure times were set to 30 -- 60 
minutes for a total integration of 6.8~hours on NGC~7789 (one exposure was 
cut short) and 5~hours on NGC~6819.  The airmass of the observations 
ranged from 1.25 -- 1.49 for the NGC~7789 spectra and from 1.07 -- 1.22 
for the NGC~6819 spectra.  For both clusters, the second priority mask was 
not observed as a 
quick reduction of the data after the first exposure indicated that 
most of the targets were in fact DA white dwarfs (e.g., broad Balmer lines 
seen in the spectra).  In total, the NGC~7789 spectroscopic field 
contained 15 targets, 9 of which were top priority white dwarf candidates.  
For NGC~6819, 13 objects were targeted in the one mask of which 8 were top 
priority candidate white dwarfs.  Additional box slits were used for 
alignment.  The locations of these 28 selected white dwarf 
candidates on the faint-blue corners of the cluster CMDs are displayed 
in Figure~\ref{fig:wdzoom}, for each of 
NGC~7789 and NGC~6819.  We have also introduced a numbering scheme to 
identify these objects later (i.e., object ``1'' in NGC~7789 is labeled 
as NGC~7789~--~1).  The selected objects sample the observed white dwarf 
cooling sequence over approximately three magnitudes, in each cluster.  
The solid curve represents a 0.6~$M_\odot$ white dwarf cooling 
sequence \citep{wood95}.  Postage-stamp cutouts of each of the 
28 white dwarf candidates from the CFHT imaging are shown in 
Figure~\ref{fig:snapshot} and the photometric properties of these 
stars are summarized in Table~2.

The spectroscopic data were analyzed as described in \cite{kalirai07}.  
Specifically, we used Python routines that are described in \cite{kelson00} 
and \cite{kelson03} to perform bias subtraction, vertical distortion 
corrections, wavelength calibration (typical $rms$ scatter in the dispersion 
solutions is $<$0.05~${\rm \AA}$), flat-field corrections, and sky subtraction.  
Standard IRAF tasks were used to extract these to 1-d spectra, co-add 
individual exposures, and flux calibrate using a spectrophotometric standard 
star (HZ~44).  Of the 28 objects targeted on the 
two masks, we recovered a spectrum for all but one.  This one faint object 
(NGC~7789~--~13) has $V$ = 23.47 and $B-V$ = 0.24 and was given a very 
short slit length as its position was between two other high priority stars.  
The multiobject data reduction for this slit failed at several steps of 
the pipeline (e.g., wavelength calibration and sky subtraction) in each 
of the individual exposures despite several attempts to recover a reduced 
spectrum.


\section{NGC 7789 and NGC 6819 White Dwarf Spectra} \label{WDspectra}


\begin{figure*}
\begin{center}
\leavevmode 
\includegraphics[height=17.0cm,angle=270]{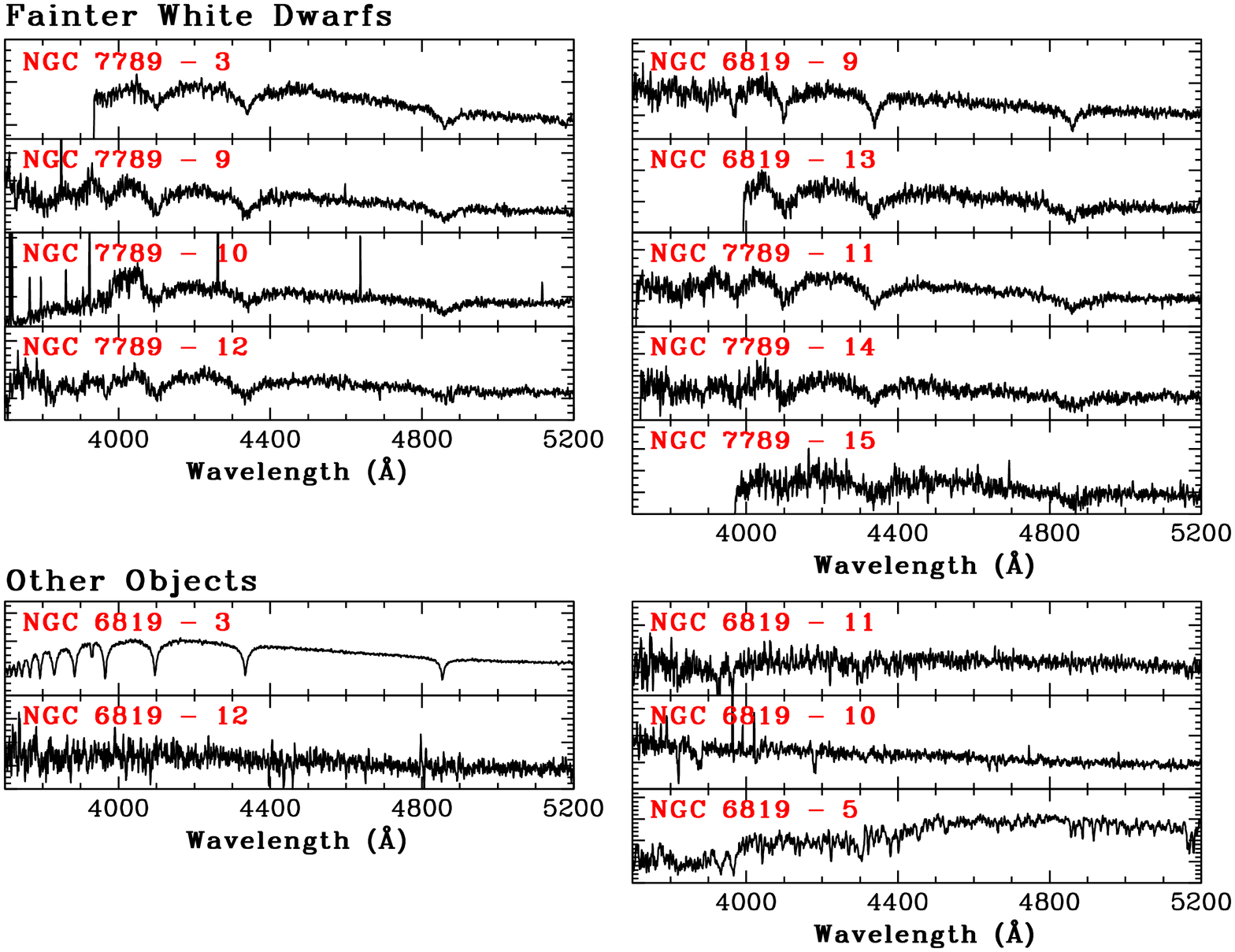}
\end{center}
\caption{{\it Top} -- Spectra for nine fainter white dwarfs in our 
data set.  These spectra are too noisy to accurately characterize the 
shapes of the higher order Balmer lines and therefore can not be used 
to yield accurate temperatures and gravities for the stars.  All of the 
stars can, however, be classified as DA white dwarfs.  {\it Bottom} -- 
Spectra for five objects of likely non-white dwarf nature as discussed 
in \S\,\ref{otherobjects}).
\label{fig:wdspectralow}}
\end{figure*}


The spectra for the 27 extracted white dwarf candidates targeted in this study 
are shown in Figures~\ref{fig:wdspectrahigh} and \ref{fig:wdspectralow}.   The 
majority of the 27 targets show clear evidence for pressure broadened Balmer 
lines and are therefore DA (hydrogen atmosphere) white dwarfs.  The top group 
of ten white dwarfs in Figure~\ref{fig:wdspectrahigh} represent our highest 
quality data, and will be used in the analysis that follows.  These 
stars are clearly among the brightest in our data set and the spectra reveal well 
defined Balmer lines from H$\beta$ down to H8 and H9.  We discuss these objects 
further in section~\ref{WDMasses}.

\subsection{Rare White Dwarfs} \label{rareobjects}

The second set of targets in Figure~\ref{fig:wdspectrahigh} (bottom) represent three 
rare objects in our sample.  Each of these stars is potentially very important 
(for different reasons) and so we discuss them in turn.  The first, object~4 in 
NGC~6819, is clearly a DB (helium atmosphere) white dwarf.  He absorption lines at 
3889, 4471, and 4713 ${\rm \AA}$ are all seen in the stellar spectrum (see three 
panels on the right in Figure~\ref{fig:wdspectrahigh} for a closer look at these 
features).  If a member of the cluster, this star therefore represents one of only 
four helium atmosphere white dwarfs found in all open clusters.  The other such 
stars are the DBA white dwarf LP~475-242 in the Hyades, the DQ white dwarf 
NGC~2168:LAWDS~28 in NGC~2168 \citep{williams06}, and the newly discovered 
DB white dwarf NGC~6633:LAWDS~16 in NGC~6633 \citep{williams07}.  Unfortunately, 
the spectral quality of NGC~6819~--~4 is too low to estimate the temperature or 
mass of the star from the helium lines.  \cite{kalirai05b} proposed that the 
absence of {\it DBs} in open clusters may be related to the fact that this population 
of white dwarfs is more massive than the field population (where we typically 
find 20--25\% DBs).  This results from the targeting of younger clusters 
in previous studies (more massive progenitor stars) that have only produced 
massive white dwarfs.  Such hot, high mass white dwarfs may not develop large enough 
helium convection zones to allow helium to be brought to the surface and turn 
a hydrogen-rich white dwarf into a helium-rich one.  Kalirai et~al.\ predicted 
that an increasing number of DB white dwarfs should be seen as observations 
begin to probe older clusters, such as NGC~6819 and NGC~7789.


\begin{figure*}
\begin{center}
\leavevmode 
\includegraphics[height=17.0cm,angle=270]{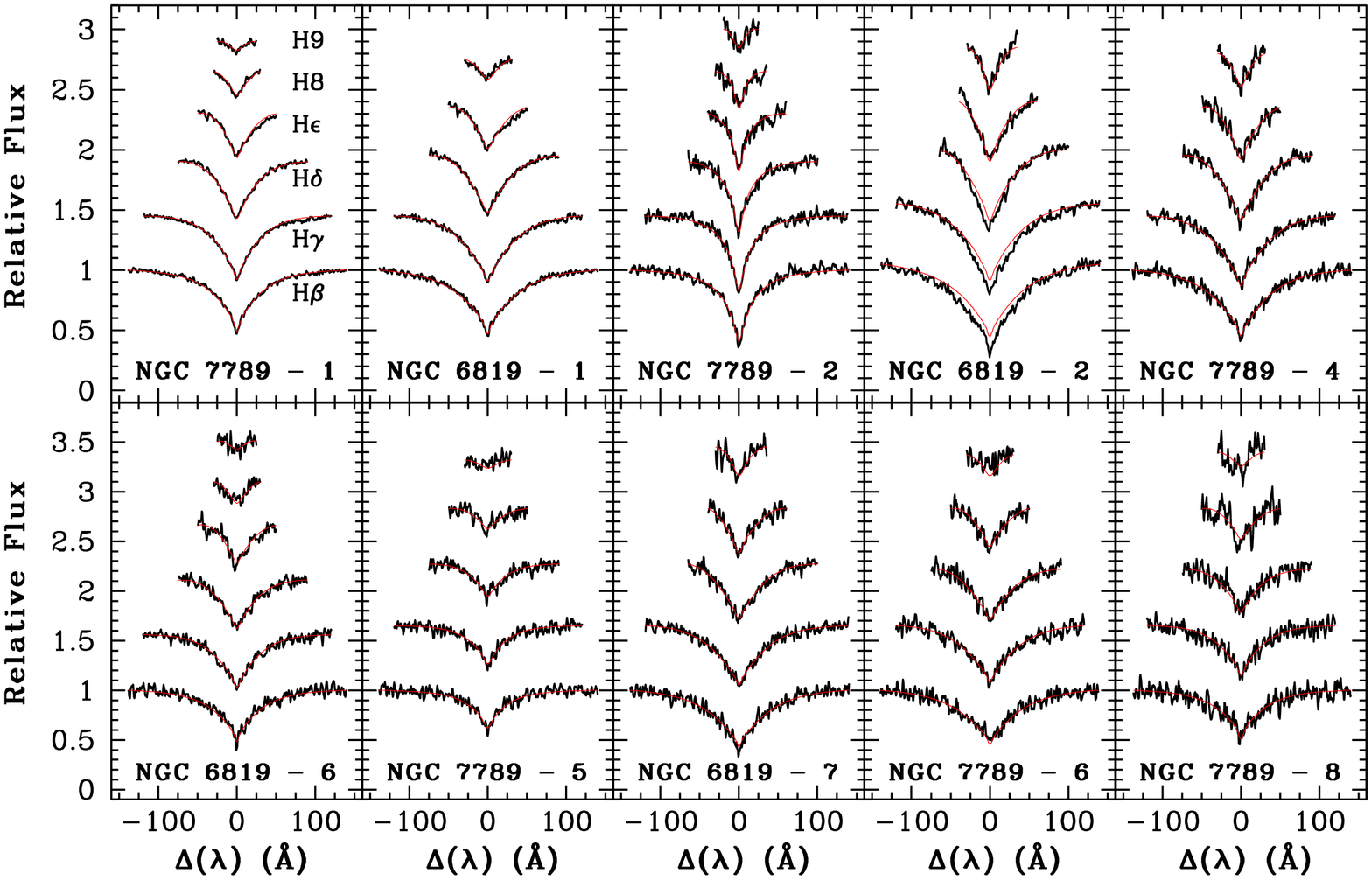}
\end{center}
\caption{The best fit hydrogen atmosphere model white dwarf 
spectrum (red curve) is shown for each of the ten stars in 
Figure~\ref{fig:wdspectrahigh} (top).  Within each of the 
panels, the hydrogen Balmer lines for a single star are arranged 
with H$\beta$ at the bottom and successively higher order lines 
towards the top.  The model shown represents the fit with the lowest 
$\chi^{2}$ to {\it all} lines simultaneously.  As discussed in 
section~\ref{WDMasses}, these fits provide accurate measurements 
of both $T_{\rm eff}$ and log~$g$ (and therefore the stellar mass) 
for each white dwarf (see Table~3).
\label{fig:wdmasses}}
\end{figure*}



The second object, NGC~6819~--~8 shows very broad H$\beta$ and H$\gamma$ absorption 
lines but an absence of higher order Balmer lines.  Although the spectral quality 
is not high enough to absolutely rule out the presence of {\it weak} higher order 
lines, this signature may suggest that the white dwarf is quite massive (see 
section~\ref{WDMasses} for more information).  Furthermore, there is evidence for 
Zeeman splitting of the lines (see two panels on the right) and therefore this object 
is likely a magnetic white dwarf (see e.g., Liebert, Bergeron, \& Holberg 2003).  
Again, the quality of the spectrum is too poor to estimate the magnetic 
field or the stellar mass and therefore it would be useful to obtain higher 
S/N spectral observations of this star.  Additionally, a reduction 
of the red side spectrum from LRIS may shed further light on this interesting 
object.  Given the poor quality, we also speculate whether the observed {\it splitting} 
may actually represent emission in the core of the Balmer absorption lines, in which 
case this object may be a binary system in which the primary is accreting material 
from the secondary star.  If the primary is massive enough, such a system could 
be a potential type Ia supernova progenitor.  \cite{hurley03} speculated that this 
cluster contains a large fraction of white dwarfs that once had binary companions, 
in addition to double degenerates, to reproduce the scatter along the cooling 
sequence.  Note, the image of this star does not show any nearby companions 
(see Figure~\ref{fig:snapshot}).

Interestingly, object 7 in NGC 7789 shows {\it both} hydrogen and helium lines 
in its spectrum.  These are again highlighted in the three panels showing different 
wavelength regions (Figure~\ref{fig:wdspectrahigh} -- bottom-right).  At lower wavelengths, 
the first panel shows hydrogen lines at 3970 and 4101 ${\rm \AA}$ (H$\epsilon$ and H$\delta$) 
as well as helium lines at 3889 and 4026 ${\rm \AA}$.  In the middle panels H$\gamma$ is 
seen at 4340 ${\rm \AA}$ as well as two helium lines at 4388 and 4471 ${\rm \AA}$.  At 
longer wavelengths (3rd panel), H$\beta$ is visible at 4861 ${\rm \AA}$ as well as 
three more helium lines at 4713, 4922, and 5016 ${\rm \AA}$.  This object(s) is therefore 
either a single DAB (mixed hydrogen and helium atmosphere) white dwarf or a double 
degenerate (i.e., unresolved white dwarf - white dwarf binary) consisting of both 
a DA and a DB (helium atmosphere) star.  Distinguishing between these two cases 
is very difficult without a much higher S/N spectrum of this target 
(see e.g., Bergeron \& Liebert 2002).  Both cases are very interesting.  DAB white 
dwarfs are rare and can shed light on diffusion 
processes in white dwarfs and help our understanding of the chemical evolution 
of these stars.  If this object is in fact a binary, then the discovery of the 
DB white dwarf would in fact represent the fifth helium atmosphere white dwarf 
in an open cluster (see above for the other four such stars).  The image of this 
source in Figure~\ref{fig:snapshot} does show two nearby neighbors however both 
of these other stars are red main sequence dwarfs and therefore can not account for 
the contaminant.

\subsection{Lower Quality White Dwarf Spectra \\ and Other Objects} \label{otherobjects}

In Figure~\ref{fig:wdspectralow} (top) we present spectra of fainter DA white dwarfs in 
our data set.  These spectra are too noisy to yield accurate spectroscopic 
masses and therefore we will ignore them in the subsequent analysis.  However, it is 
reassuring that most of the faint blue targets in our spectroscopic study are in fact 
white dwarfs.  First, this suggests that our target selection process in these rich 
clusters is efficient.  As mentioned earlier, we also found a high success rate in our 
study of NGC~2099 \citep{kalirai05a} and NGC~6791 \citep{kalirai07}.  Second, these 
fainter white dwarfs can be followed up with future observations to improve the 
S/N of the spectra and therefore may eventually be important in placing 
constraints on the initial-final mass relation.  The faintest white dwarfs may even 
represent descendents from more massive main-sequence stars and therefore allow a 
probe of the relation over a mass range, within a given cluster.  

At the bottom of Figure~\ref{fig:wdspectralow} are spectra for five other objects along 
our line of sight.  The most interesting case is NGC~6819~--~3 which exhibits the 
hydrogen Balmer series although the lines are not pressure broadened.  This object 
is therefore either a field horizontal branch star, or a distant main-sequence 
dwarf of spectral type late A or early F.  If the latter, the observed magnitude 
of the star ($V$ = 21.9) implies a distance of $\sim$80~kpc.  Similarly, 
NGC~6819~--~11 and 5 appear to be background dwarfs of later spectral type.  The 
other two spectra lack enough signal to accurately classify the objects.  These could 
be cool DA white dwarfs, DB or DC white dwarfs, or other objects along the line of sight 
such as unresolved blue galaxies.

Overall, our spectra confirm that 22 of the 27 targets for which we extracted a 
spectrum are in fact white dwarf stars.  We now proceed to analyze further the 
six white dwarfs in NGC~7789 and four white dwarfs in NGC~6819 that show well 
characterized Balmer lines (i.e., the top group in Figure~\ref{fig:wdspectrahigh}).


\section{The Masses of White Dwarfs in \\ NGC~7789 and NGC~6819} \label{WDMasses}

Several techniques exist to measure the masses of white dwarfs, depending on 
what information is available.  If the star is in a binary system, a 
dynamical mass estimate can be easily calculated from the orbit of the two 
stars.  For example, the nearest white dwarf Sirius~B was known to exist 
as early as 1841 from its dynamical influences on the optically brighter 
companion, Sirius A (Bessel~1844).  The optical detection of the white dwarf 
did not occur until 1862 (by Alvan Clark), shortly after which the star was 
known to be a $\sim$1~solar mass object from the period of the binary 
($\sim$50~years).  For a white dwarf with a known radial velocity, 
the gravitational redshift method can also be used to measure the stellar mass 
(e.g., Adams~1925; Wegner~1989; Reid~1996).  Given the large gravity, photons from 
the surface of the white dwarf will lose energy as they escape the potential 
of the star and therefore be redshifted (as first suggested by Michell 1784).  
To measure this effect, the H$\alpha$ Balmer line at 6563~${\rm \AA}$ is typically 
observed at intermediate resolution.  Other methods to measure white dwarf 
masses are applicable to smaller subsets of stars only, e.g., pulsation 
mode analysis of very hot white dwarfs \citep{kawaler91} and fits to the 
mass-radius relation for stars with trigonometric parallaxes \citep{koester79}.

The most widely adopted technique for measuring the mass of a white dwarf 
involves fitting the Balmer lines of the spectrum to model atmospheres 
\citep{bergeron92}.  The shapes of these line profiles depends sensitively on 
changes in the temperature ($T_{\rm eff}$) and surface gravity (log~$g$) of the 
star.  For example, as the atmospheric pressure in a white dwarf 
increases (e.g., due to a larger surface gravity), interactions between 
neighboring hydrogen atoms will lead to enhanced Stark broadening.  For 
the lower order Balmer lines, this means the profiles will become broader 
(e.g., H$\beta$ and H$\gamma$).  However, the bluer Balmer lines are 
produced by electron transitions at higher energy levels and therefore 
these lines will be the first to be destroyed by the increased perturbations 
on the atom (e.g., H$\epsilon$, H8, H9, etc...).  As an example of this, 
see Figure~3 in \cite{bergeron92}.  Therefore, using this technique to 
accurately define the $T_{\rm eff}$ and log~$g$ of a white dwarf requires 
the characterization of higher order Balmer lines in the stellar spectrum.  
For faint stars, this implies the need for a blue-sensitive spectrograph as
the wavelength of H$\epsilon$ is in the violet region, 3970 ${\rm \AA}$.  
With $T_{\rm eff}$ and log~$g$ constrained, the mass of the star can be 
obtained through a mass-radius relation.

A significant sample of white dwarfs has been observed using at least two of 
these techniques (including the spectroscopic Balmer line fitting technique) 
and therefore provide a means to independently check the accuracy of the 
method.  \cite{bergeron95} analyze 35 such white dwarfs and find 
a reasonable agreement between spectroscopic mass determinations and 
gravitational redshifts for only those stars with $T_{\rm eff} >$ 
12,000~K.  For the cooler stars, the spectroscopic mass determinations are 
systematically larger than the gravitational redshifts by $\sim$0.1~$M_\odot$.  
As pointed out by \cite{bergeron95}, these measurements could be 
in error if convection has set in and polluted the atmospheres of 
these cool stars with helium (i.e., this would mimic a larger mass).  A similar 
study by \cite{reid96} based on HIRES spectra of 53 white dwarfs also found 
good agreement between these two methods for white dwarfs with 
$T_{\rm eff} >$ 14,000~K (see also Claver et~al.\ 2001).  Finally, a recent 
study with {\it HST}/STIS has made it possible to calculate the mass of Sirius~B 
using three independent techniques \citep{barstow05}.  The mass of the white 
dwarf based on its orbit, gravitational redshift, and blue Balmer lines, all 
indicate that the star is one solar mass to within a few percent.  As we show below, 
all but one of our white dwarfs have $T_{\rm eff} >$ 13,000~K and therefore the 
spectroscopic mass measurements are not affected by any of these possible 
systematic errors.

The fitting technique to derive $T_{\rm eff}$ and log~$g$ is described 
in \cite{bergeron92}.  We convolved the models with 
a Gaussian profile with FWHM = 4~${\rm \AA}$ to match the resolution of 
our spectra.   All of the available Balmer lines of each star are fit 
simultaneously and the best fit solution is converged upon by minimizing 
$\chi^{2}$ using the nonlinear least-squares method of Levenberg-Marquardt 
\citep{press86}.  In this fit, the estimation of the continuum near each 
Balmer line is performed using the upgraded method described in 
\cite{liebert05}.  The atmosphere models cover a log~$g$ range from 
6.5 -- 9.0 and a $T_{\rm eff}$ range from 1500 -- 100,000~K.  The best 
solutions for the ten white dwarfs in NGC~7789 and NGC~6819 are illustrated 
in Figure~\ref{fig:wdmasses}.  For each star, we present the observed Balmer lines one 
on top of another, with H$\beta$ at the bottom and subsequent higher order 
lines at the top (up to H9 at 3835~${\rm \AA}$).  The 
best fit model solution for each is shown as a smooth profile (red curve). 
The fits are excellent in all cases except for NGC~6819~--~2 in which the 
lower order Balmer lines are not reproduced as well as the higher order 
lines.  If we ignore the higher order lines of this star and refit 
only H$\beta$, H$\gamma$, and H$\delta$, the quality of the fit does 
not improve and the derived parameters of the star remain essentially 
unchanged.  As we below show in section~\ref{membership}, this star is 
{\it not} a cluster member and therefore does not enter into our 
analysis of the initial-final mass relation.  

White dwarf masses ($M_{\rm final}$) are calculated for each star by interpolating 
the $T_{\rm eff}$ and log~$g$ within the updated evolutionary models of 
\cite{fontaine01} for a 50/50 carbon-oxygen core mix.  The models adopt thick 
hydrogen layers ($q(\rm H)$ = $M_{\rm H}/M$ = 10$^{-4}$) and helium layers of 
$q(\rm He)$ = 10$^{-2}$.  The models also provide white dwarf cooling ages ($t_{\rm 
cool}$) for each star (i.e., the age of the star since shell helium burning finished 
on the asymptotic giant branch).  We summarize the spectroscopic properties for 
these twelve white dwarfs in Table~3.  


\section{Calculating Main-Sequence Progenitor Lifetimes and Masses} \label{MSlifetimes}

Unlike for the field population of isolated white dwarfs, the environments of 
white dwarfs in star clusters can be used to shed light on the properties of 
their progenitors.  As star clusters are co-eval, the main-sequence turnoff 
ages of NGC~7789 and NGC~6819 measured in section~\ref{red.dist.age} also 
represent the {\it total} lifetime of their inhabiting white dwarfs (i.e., 
the main-sequence lifetime plus the timescales for evolutionary stages beyond 
core hydrogen burning).  Therefore, by subtracting the white dwarf cooling age 
from the cluster age, we can calculate the lifetime of the progenitor star that 
made the white dwarf up to the tip of the asymptotic giant branch.  For clusters 
as old as NGC~7789 and NGC~6819, this latter age ($t_{\rm ms}$) is dominated by 
the main-sequence lifetime of the star since the post main-sequence evolutionary 
phases are short lived.

\subsection{Cluster Membership} \label{membership}

The assumption in the above calculation is that the spectroscopically confirmed 
white dwarfs in this study are in fact members of NGC~7789 and NGC~6819.  As the 
volume probed increases with photometric depth, most of the field white dwarfs 
along these lines of sight will be found near the faint end of the data set.  
Depending on the distance and age of any field white dwarf, it could mimic 
itself as a faint-blue cluster object.

\begin{figure}
\epsscale{1.1} \plotone{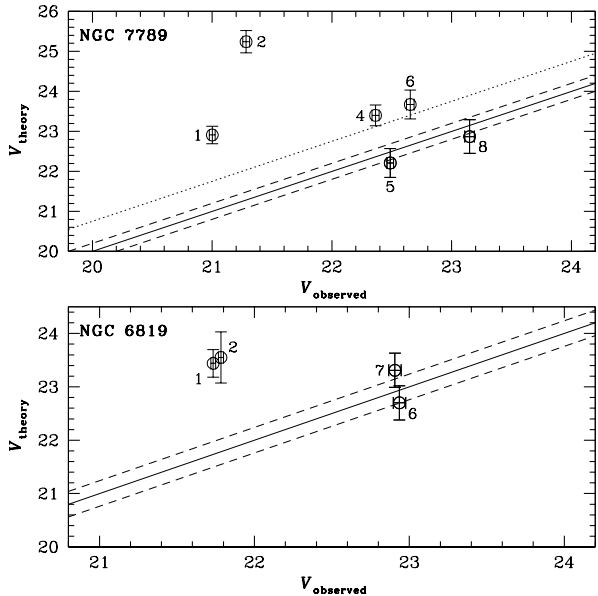} 
\figcaption{A comparison of the theoretical magnitude of the star (from fitting 
the Balmer lines) with the observed brightness indicates that two of the 
white dwarfs in both NGC~7789 and NGC~6819 are single white dwarfs (darker objects) 
in these clusters (2$\sigma$ error bars).  An additional two objects in NGC~7789, 
objects 4 and 6, are consistent with being 0.75 magnitudes overluminous and 
therefore may represent unresolved double degenerate systems.
\label{fig:mags}}
\end{figure}

To determine which of the stars are likely cluster white dwarfs, 
we use the white dwarf mass-radius relation to calculate a theoretical 
magnitude for each star.  This magnitude is next compared to the observed brightness 
of the respective white dwarf by adopting the distance modulus of each 
cluster derived in section~\ref{red.dist.age} (($m-M$)$_V$ = 12.5 $\pm$ 
0.1 for NGC~7789 and ($m-M$)$_V$ = 12.30 $\pm$ 0.12 for NGC~6819).
Figure~\ref{fig:mags} shows the results.  The solid line represents the 
1:1 relation and the dashed lines are 2$\sigma$ bounds based on the 
distance errors above.  The uncertainties on the data points are also 
2$\sigma$ error bars.  In NGC~7789, a group of four white dwarfs are 
found near the 1:1 relation, and two others are obvious outliers 
(objects 1 and 2).  Based on this diagram, only objects 5 and 8 can be 
considered {\it isolated} cluster members (darker points).  However, we 
note that both objects 4 and 6 are consistent with a 0.75 magnitude offset 
from the 1:1 relation (observed magnitude being too bright -- dotted 
line).  These stars are therefore overluminous by an amount consistent 
with an equal mass binary nature (i.e., they could be unresolved double 
degenerates in the cluster), assuming they are not optical binaries.  For 
NGC~6819, Figure~\ref{fig:mags} indicates that objects 6 and 7 are cluster 
members whereas objects 1 and 2 are classified as non members based 
on our parameter measurements.


For the four non-binary cluster member stars and the two potential double degenerate 
systems, we measure the main-sequence plus post main-sequence lifetimes (up to the 
tip of the asymptotic giant branch) by subtracting the derived white dwarf cooling 
ages from the cluster ages ($t$ = 1.4~Gyr for NGC~7789 and $t$ = 2.5~Gyr for NGC~6819 
-- see section~\ref{red.dist.age}).  These results, $t_{\rm ms}$, are given in 
column~7 of Table 3.  The main-sequence masses ($M_{\rm initial}$) follow from the 
models of \cite{hurley00} and are listed in column~8 of Table~3.  The errors in the 
main-sequence lifetimes include the uncertainties in the cooling ages and an assumed 
10\% uncertainty in the ages of the clusters.


\begin{figure}
\epsscale{1.1} \plotone{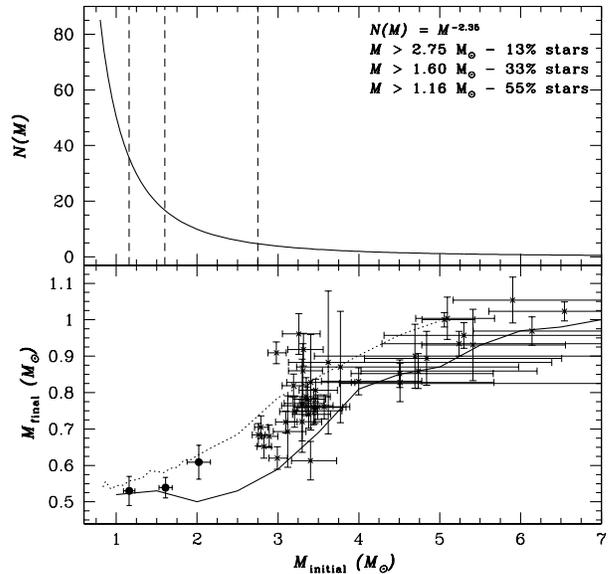} 
\figcaption{{\it Top} -- For a Salpeter mass function only 13\% of 
all stars are formed with $M >$ 2.75~$M_\odot$ whereas 55\% have 
$M >$ 1.16~$M_\odot$. {\it Bottom} -- All previous constraints on the 
initial final mass relation (crosses -- see references in \S\,\ref{lowmassend}) 
and weighted averages for the stars in this study (filled circles) 
from NGC~7789 ($M_{\rm initial}$ = 2.03~$M_\odot$) and NGC~6819 
($M_{\rm initial}$ = 1.61~$M_\odot$), as well as the masses of the 
single carbon-oxygen core white dwarf and progenitor in NGC~6791 
($M_{\rm initial}$ = 1.16~$M_\odot$).  The uncertainties on these three 
points represent 2$\sigma$ errors.  The new data extend the 
initial-final mass relation to very low masses and show that 
the observed trend at higher masses continues down to stars approximately 
the mass of the Sun.  The two curves are the core mass at the first 
thermal pulse (solid) and the solar metallicity theoretical initial 
final mass relation (dotted) from Marigo (2001).
\label{fig:ifmr}}
\end{figure}



\begin{table*}
\begin{center}
\caption{}
\begin{tabular}{lcccclccc}
\hline
\hline
\multicolumn{1}{c}{ID} & 
\multicolumn{1}{c}{$V_{\rm theory}$$^{a}$} & \multicolumn{1}{c}{$T_{\rm eff}$ (K)} & 
\multicolumn{1}{c}{log~$g$} & \multicolumn{1}{c}{$M_{\rm final}$ ($M_\odot$)} & 
\multicolumn{1}{c}{$t_{\rm cool}$ (Myr)} & \multicolumn{1}{c}{$t_{\rm ms}$ (Myr)} & 
\multicolumn{1}{c}{$M_{\rm initial}$ ($M_\odot$)} & \multicolumn{1}{c}{Member?} \\
\hline
NGC 7789~--~1       &  22.91 $\pm$ 0.11 & 21,900 $\pm$ 100  & 7.89 $\pm$ 0.01 & 0.56 $\pm$ 0.01 & 35   $\pm$  1 & ------------ & ------------             & no \\
NGC 7789~--~2       &  25.24 $\pm$ 0.14 &  9,700 $\pm$  50  & 8.31 $\pm$ 0.03 & 0.80 $\pm$ 0.02 & 1040 $\pm$ 64 & ------------ & ------------             & no \\
NGC 7789~--~4$^{b}$ &  23.40 $\pm$ 0.13 & 16,900 $\pm$ 200  & 7.90 $\pm$ 0.03 & 0.56 $\pm$ 0.02 & 115  $\pm$  8 & 1285 $\pm$ 140 & $2.08^{+0.08}_{-0.08}$ & ? \\ 
NGC 7789~--~5       &  22.21 $\pm$ 0.18 & 31,200 $\pm$ 200  & 7.90 $\pm$ 0.05 & 0.60 $\pm$ 0.03 & 8    $\pm$  1 & 1392 $\pm$ 140 & $2.02^{+0.07}_{-0.14}$ & yes \\
NGC 7789~--~6$^{b}$ &  23.67 $\pm$ 0.18 & 17,600 $\pm$ 300  & 8.15 $\pm$ 0.06 & 0.72 $\pm$ 0.03 & 160  $\pm$ 16 & 1240 $\pm$ 141 & $2.10^{+0.09}_{-0.09}$ & ? \\ 
NGC 7789~--~8       &  22.87 $\pm$ 0.21 & 24,300 $\pm$ 400  & 8.00 $\pm$ 0.07 & 0.64 $\pm$ 0.04 & 29   $\pm$  5 & 1371 $\pm$ 140 & $2.02^{+0.09}_{-0.11}$ & yes \\
NGC 6819~--~1       &  23.44 $\pm$ 0.13 & 19,600 $\pm$ 100  & 8.25 $\pm$ 0.02 & 0.78 $\pm$ 0.01 & 130  $\pm$  5 & ------------ & ------------             & no  \\  
NGC 6819~--~2       &  23.55 $\pm$ 0.24 & 13,100 $\pm$ 600  & 7.82 $\pm$ 0.07 & 0.50 $\pm$ 0.04 & 261  $\pm$ 36 & ------------ & ------------             & no  \\ 
NGC 6819~--~6       &  22.70 $\pm$ 0.16 & 21,100 $\pm$ 300  & 7.83 $\pm$ 0.04 & 0.53 $\pm$ 0.02 & 39   $\pm$  3 & 2461 $\pm$ 250 & $1.60^{+0.06}_{-0.05}$ & yes \\
NGC 6819~--~7       &  23.31 $\pm$ 0.16 & 16,000 $\pm$ 200  & 7.91 $\pm$ 0.04 & 0.56 $\pm$ 0.02 & 143  $\pm$ 11 & 2357 $\pm$ 250 & $1.62^{+0.07}_{-0.05}$ & yes \\
\hline
\end{tabular}
\tablenotetext{$^a$}{Theoretical luminosity from spectral fits (see \S\,\ref{membership}).}
\tablenotetext{$^b$}{Possible cluster double degenerates.}
\label{table3}
\end{center}
\end{table*}



\section{The Initial-Final Mass Relation} \label{ifmr}

\subsection{Constraining the Low Mass End} \label{lowmassend}

We stressed earlier the importance of a well constrained initial-final 
mass relation {\it over a wide mass range}.  Star formation in the 
Universe leads to an initial mass function that is generally steep 
(i.e., many more low mass stars are produced as compared to high mass 
stars -- Salpeter~1955; Miller \& Scalo 1979; Kroupa 2002).  
We illustrate a simple mass function with a Salpeter 
slope in Figure~\ref{fig:ifmr} (top), for 1000 stars over a 
mass range of 0.8 -- 7 $M_\odot$.  These limits have been chosen as 
they range from the lowest mass stars that could have formed white dwarfs 
over the age of the Universe ($\sim$0.8~$M_\odot$) to the most massive 
such stars ($\sim$7~$M_\odot$).  In the bottom panel, we illustrate 
the initial-final mass relation with all constraints over the past 
30 years (crosses).  This includes white dwarfs in the Hyades, Praesepe, 
and Pleiades clusters \citep{claver01,dobbie04,dobbie06}, NGC~3532 
\citep{koester93}, NGC~2516 \citep{koester96}, NGC~2168 \citep{williams04}, 
NGC~2099 \citep{kalirai05a}, NGC~6633 \citep{williams07}, and Sirius~B 
\citep{liebert05b}.  Initial and final masses are taken from Table~1 in 
Ferrario et~al.\ (2005).  The only stars ignored in this analysis are 
four white dwarfs in young clusters with masses $<$0.55~$M_\odot$ that 
likely represent field contamination (e.g., see Kalirai et~al.\ 2005a 
for NGC~2099) and two stars with $>$90\% uncertainties in their initial 
masses (star 3532-10 in NGC~3532 and 2099-WD16 in NGC~2099).  Over the region 
where information is available ($M_{\rm initial} >$ 
2.75 $M_\odot$ -- right dashed line), the relation shows a trend 
indicating that more massive main-sequence stars produce more 
massive white dwarfs.  Integrating the Salpeter mass function above 
this limit, we find that {\it only} 13\% of all stars are born with 
masses this large over the 0.8 -- 7 $M_\odot$ mass range.  Therefore, 
the present initial-final mass relation can not be directly used to 
infer progenitor properties for almost all white dwarfs in the Galactic 
disk and halo.

The 2.75 $M_\odot$ lower initial mass limit on the relation results 
purely from an observational limitation.  A low mass (0.6 $M_\odot$), 
bright white dwarf has $M_V \sim$ 11 at an age of $\sim$100 Myr.  At 
a distance of 1.5~kpc, this translates to an observed magnitude of 
$V \sim$ 22.  A more massive white dwarf at this age will be even 
fainter in the $V$ band.  As discussed above, measuring a spectroscopic mass 
for a white dwarf requires the accurate characterization of higher order 
Balmer lines with $\lambda <$ 4000 ${\rm \AA}$.  Achieving this for 
a $V$ = 22 star obviously requires both a large telescope and a blue-sensitive 
spectrograph.  Since very few rich (e.g., $>$1000~$M_\odot$), old star 
clusters are located within 1.5~kpc of the Sun (M67 
is the only one), the targeted systems have typically been poorly 
populated, nearby, younger systems (such as the Hyades, Pleiades, 
and Praesepe clusters).  These clusters have ages of a few 
hundred Myr and therefore are not old enough to have allowed the 
evolution of lower mass stars off the main sequence.

The masses of white dwarfs in systems such as NGC~7789 and NGC~6819 
present us the opportunity to extend the mass range over which the 
initial-final mass relation has been studied.  The younger of our 
two clusters, NGC~7789, has an age of 1.4~Gyr.  The main-sequence 
turnoff of this system is therefore 2.0~$M_\odot$ \citep{vandenberg06} 
and most of the cluster white dwarfs will have evolved from stars just 
above this mass (again, due to the slope of the mass function).  
The age of NGC~6819 is 2.5~Gyr and therefore the present day turnoff mass 
is 1.6~$M_\odot$.  The progenitor masses for the confirmed white dwarfs 
in each cluster are indeed very similar to one another, and just above 
the turnoff masses (see $M_{\rm initial}$ in Table~3).  

We can also add data from the very old cluster NGC~6791 to the initial-final 
mass relation.  At an age of 8.5~Gyr, this system represents one of the oldest 
open star clusters and has a main-sequence turnoff mass of 
$\sim$1.1~$M_\odot$.  \cite{kalirai07} present evidence that a significant population 
of white dwarfs in this cluster resulted from progenitors that expelled enough 
mass on the red giant branch to avoid the helium flash, and therefore the white 
dwarfs have helium cores rather than carbon-oxygen cores.  This is believed to 
be a result of the high metallicity of the system, [Fe/H] = $+$0.4.  
The mean mass of the nine cluster white dwarfs targeted in that study is 
$\langle$$M$$\rangle$ = 0.43~$M_\odot$.  The threshold at which a helium-core white 
dwarf is produced at NGC~6791's metallicity is 0.45 -- 0.47 $M_\odot$, and 
therefore to be conservative, we consider only the single confirmed cluster member 
with a mass $>$0.50~$M_\odot$ (definite carbon-oxygen core white dwarf).  This 
object, NGC~6791 WD~7, has $M_{\rm initial}$ = $1.16^{+0.04}_{-0.03}$ and 
$M_{\rm final}$ = 0.53 $\pm$ 0.02~$M_\odot$ (see Kalirai et~al.\ 2007 for the 
spectral fits).  The initial mass for this star has been calculated using the 
same \cite{hurley00} models as for NGC~7789 and NGC~6819, for $Z$ = 0.035 (the 
highest metallicity available in these models).  We note that this data point 
may still represent a lower limit (i.e., the final mass) since the progenitor 
star of the carbon-oxygen core white dwarf also likely suffered from enhanced 
mass loss.  The weighted mean progenitor mass and white dwarf mass for the two 
stars in NGC~7789, the two stars in NGC~6819, and the masses of the single 
object in NGC~6791 are

\begin{eqnarray*}
M_{\rm initial} = 2.02 \pm 0.07~M_\odot,  M_{\rm final} = 0.61 \pm 0.02~M_\odot \\
M_{\rm initial} = 1.61 \pm 0.04~M_\odot,  M_{\rm final} = 0.54 \pm 0.01~M_\odot \\
M_{\rm initial} = 1.16 \pm 0.04~M_\odot,  M_{\rm final} = 0.53 \pm 0.02~M_\odot.
\end{eqnarray*}

\noindent We illustrate these new points as filled circles on the initial-final 
mass relation in Figure~\ref{fig:ifmr} (bottom).  The calculated initial 
progenitor masses (for bright white dwarfs) in these much older clusters are all 
essentially the same as the cooling ages are a very small fraction 
of the cluster ages.  We have therefore plotted one data point for each 
cluster, which for the two clusters with multiple white dwarfs, represents the 
weighted mean of the system's progenitor and white dwarf masses.  Also shown is 
the 2$\sigma$ error in each quantity for all three clusters (see 
section~\ref{parameterizing} for more information on this).  As expected, 
these new data points provide constraints on the low mass end of the relation.  
They clearly indicate that the observed trend at higher masses (suggesting more 
massive main-sequence stars produce more massive white dwarfs) continues down 
to stars that are roughly one solar mass.

\subsection{Theoretical Estimates of Stellar Mass Loss} \label{massloss}

Most of the mass loss that a star suffers through its evolution occurs during 
very short lived post-main-sequence evolutionary phases such as the red giant 
branch, asymptotic giant branch, and planetary nebula phases (e.g., see Reimers~1975).  
In fact, it is the mass loss that is responsible for concluding fusion processes in the 
star and hence its rise in luminosity on the asymptotic giant branch.  In principal, 
the masses of the stellar cores during these last phases of stellar evolution can 
be determined directly from modeling these evolutionary stages.  An ideal model 
would then take an initial star of a certain mass ($M$ $\lesssim$ 8~$M_\odot$) 
and propagate it through all phases of stellar evolution to yield a remnant 
white dwarf with a particular mass.  In practice, this has been very difficult because 
the mass loss mechanisms (e.g., helium flash and thermal pulses on the asymptotic 
giant branch) are not theoretically understood well enough (Weidemann~2000; also 
see Habing 1996 for a review).  Direct observational constraints are rare given 
the very short lifetimes of stars on the asymptotic giant branch and planetary 
nebula phases ($\sim$10$^5$ years), and heavy obscuration of sources by dusty shells.

Recently, a few attempts have been made to calculate the rate of mass loss in 
asymptotic giant branch stars after factoring in parameters such as 
metallicity (e.g., Marigo~2001).  We first present the mass of the stellar 
core at the first thermal pulse from one such calculation as a solid line in the 
bottom panel of Figure~\ref{fig:ifmr} \citep{girardi00}.  As expected, this curve 
falls below the bulk of the data points as it represents an evolutionary point 
before the core of the star has had a chance to grow during the thermal pulses.  
Note that the mass of the core is roughly constant for masses below 2~$M_\odot$.  
From this initial point, \cite{marigo01} performs synthetic calculations 
of the subsequent thermally pulsating phases of the asymptotic giant branch 
until the star has completely ejected its envelope (see details in her 
paper).  This can therefore be used to predict both the total mass loss and specific 
chemical yields as a function of initial mass and metallicity.  For solar 
metallicity, the theoretical initial-final mass relation from this work 
is shown as the dotted line in Figure~\ref{fig:ifmr} (bottom).  For 
$M_{\rm initial} >$ 4~$M_\odot$,  this curve is systematically higher 
than the observed data, predicting final remnant masses that are too 
large by up to 0.1~$M_\odot$.  A test of this relation at the low mass end 
(i.e., the new data points with $M \lesssim$ 2~$M_\odot$) also finds 
final masses that are larger than our observations, however, the differences 
are very small.  Part of this difference may even be expected in the case 
of NGC~6791 given the 2.5$\times$ higher metallicity of this cluster as compared 
to the solar metallicity theoretical relation (see earlier discussion).


\begin{figure}
\epsscale{1.1} \plotone{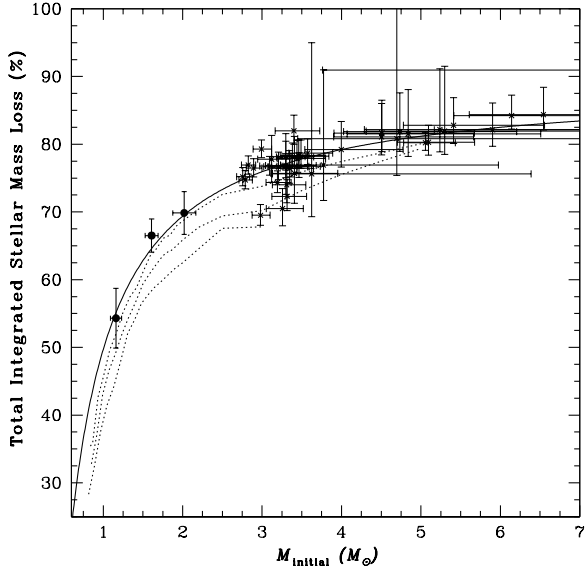} 
\figcaption{The total integrated mass loss is found to decrease as 
the initial mass of the star decreases.  The new data points from 
this study, with 2$\sigma$ error bars, are shown as filled circles.  
The dotted lines are theoretical calculations from \cite{marigo01} 
for solar metallicity (top), $Z$ = 0.008 (middle), and 
$Z$ = 0.004 (bottom).  The solid curve is the best fit linear 
least squares parameterization of the initial-final mass 
relation from this work (see section~\ref{parameterizing}).
\label{fig:massloss}}
\end{figure}


In Figure~\ref{fig:massloss} we present a different view of the initial-final 
mass relation to highlight the desired output from this work.  The vertical 
axis now shows the total integrated mass loss through stellar evolution.  For 
the most massive main-sequence stars that will form white dwarfs, this yield 
is about $\sim$85\% (e.g., the progenitor of white dwarf LB1497 in the Pleiades 
cluster).  A slightly less massive star such as the progenitor of Sirius~B 
(5.06~$M_\odot$ -- Liebert et~al.\ 2005) has lost 80\% of its mass.  The mass 
loss smoothly decreases with stellar mass down to $\sim$75\% for intermediate 
mass stars, 3 $<$ $M_{\rm initial} <$ 4 $M_\odot$.  Our new data points suggest a more 
rapid decline for stars with $M \lesssim$ 2~$M_\odot$.  At this mass, stars 
will lose $\sim$70\% of their total mass however this decreases down to just 
$\sim$55\% for stars approximately the mass of the Sun.  The theoretical 
calculation for solar metallicity discussed above is shown as the uppermost 
dotted curve \citep{marigo01}.

\subsection{The Scatter in the Relation} \label{scatter}

Several authors have commented on the observed scatter in the initial-final 
mass relation (e.g., Ferrario et~al.\ 2005).  The present data set is 
very heterogeneous.  The points on Figure~\ref{fig:ifmr} are derived from 
white dwarf observations in over ten star clusters.  The quality of these 
data and procedures used to fit the spectra vary from one investigation to 
another and therefore small biases are likely to exist in the $M_{\rm final}$ 
values.  A small amount of field contamination may even exist in the 
sample.  Additionally, the ages of the star clusters have been derived by 
different authors using different assumptions, techniques, and isochrones 
and therefore the calculations involved in determining $M_{\rm initial}$ will 
also have biases.  Even within an individual study, the large error bars in 
$M_{\rm initial}$ for the massive stars in Figure~\ref{fig:ifmr} are a good 
example of the difficulty in assigning masses to main-sequence lifetimes 
(see Ferrario et~al.\ 2005 for a version of the relation with stars from 
individual clusters color coded).  A small shift in the age of a cluster 
from 80 to 100~Myr results in a $>$0.5~$M_\odot$ systematic change in the 
inferred main-sequence mass at the turnoff.  
Measuring the ages of clusters to this precision is very difficult for 
such young systems where the morphology of the turnoff is essentially 
vertical on an optical CMD.  This is of course not a large concern in the 
study of older clusters since the turnoff can be well defined and the turnoff 
mass does not sensitively depend on the age.


\begin{figure}
\epsscale{1.1} \plotone{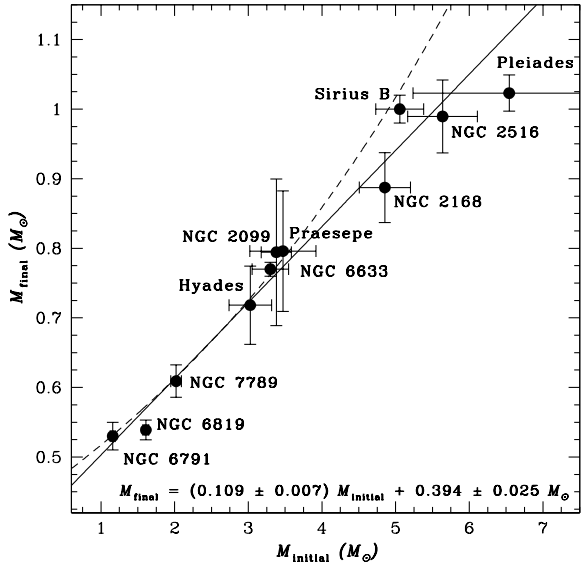} 
\figcaption{An initial-final mass relation is constructed by showing 
all of the stars from each cluster as a single data point, as labeled.  
The best-fit linear least squares relation (solid curve) is indicated 
in the panel, and is found to provide an adequate fit to the data 
(reduced chi-squared per degree of freedom is $\chi^2$ = 1.2).  The 
dashed curve shows the initial-final mass relation calculated by 
\cite{hansen07} to fit the white dwarf cooling sequence of the globular 
cluster NGC~6397 (see section~\ref{conclusion}).
\label{fig:ifmrcluster}}
\end{figure}


Although characterizing the errors resulting from these uncertainties and 
heterogeneities is difficult, it is important to distinguish these biases 
from {\it intrinsic} scatter that may result from fundamental properties of 
stellar evolution.  One way to minimize the systematic effects is to limit 
the study of a particular 
question to just the constraints from a few star clusters with many white 
dwarfs.  For example, the Hyades cluster \citep{perryman98} is of similar age 
to NGC~2099 (Kalirai et~al.\ 2001c; 2005a) yet its chemical abundance is 
enriched by a factor of two ($Z_{\rm Hyades}$ = 0.025 and $Z_{\rm NGC~2099}$ 
= 0.013).  Models of stellar evolution predict that stars of higher 
metallicity will lose mass in post main-sequence phases more efficiently 
than stars of lower metallicity (e.g., Marigo 2001).  This is illustrated in 
Figure~\ref{fig:massloss}.  As we said above, the dotted line at the top 
represents the theoretical estimates for mass loss in solar metallicity 
stars.  The two dotted lines underneath are the same relation for more 
metal-poor stars, $Z$ = 0.008 and $Z$ = 0.004 \citep{marigo01}.  Fortunately, 
both the Hyades and NGC~2099 harbor significant white dwarf populations that have been 
studied spectroscopically.  \cite{kalirai05a} showed that the mean mass 
of the NGC~2099 white dwarf population appears to be more massive 
(by $\sim$10\%) than the Hyades stars, qualitatively consistent with the 
expectations from stellar evolution (this is a 2$\sigma$ effect in the 
mean mass of the populations).  As already discussed, a convincing 
example of the efficiency of mass loss on metallicity is presented in 
\cite{kalirai07}.  These cases highlight how the different properties of 
stars may play a role in contributing to the observed scatter on the 
initial-final mass relation.  

Other properties of stars may also be important in understanding the scatter 
in the relation, such as rotation, binary evolution, and magnetic fields 
(see Weidemann~2000).  Unfortunately, the quality of the present data does 
not permit a study of these effects.  The mass loss mechanisms may themselves 
be stochastic to some degree.  For example, \cite{reid96} measured masses of 
white dwarfs in the Praesepe cluster using gravitational redshifts and 
found a large dispersion in the remnant mass distribution (0.6 -- 0.9~$M_\odot$).  
An estimate of the initial masses of these stars suggests that they were 
all produced from stars of about the same mass.  This would then suggest 
that there is no singular initial-final mass relation.  However, \cite{claver01} 
reconcile this picture by suggesting that one of the outlier stars in the 
$M_{\rm initial}$ vs. $M_{\rm final}$ plane of the Praesepe sample (LB~5893) 
may have formed from close binary evolution.  A better understanding of these 
types of effects will require a larger data set, as we discuss below.

\bigskip

\subsection{Semi-Empirical Relations and Parameterization} \label{parameterizing}

Deriving a functional form of the initial-final mass relation from the available 
data is problematic for several reasons.  First, for the reasons discussed above, 
such a parameterization may be meaningless given the uncertain degree to which 
second order properties of stars may effect their mass loss.  Second, the relation 
is only constrained over a fraction of the total mass range that is of interest.  
Prior to this work, the low mass end of the relation was completely devoid of any 
observations of individual white dwarfs with direct mass measurements.  The high 
mass end continues to be sparsely populated, the degree to which depends on the 
maximum mass of a star that will form a white dwarf (the current high mass point 
is at $M_{\rm initial}$ = 6.5~$M_\odot$, Ferrario et~al.\ 2005). 

\cite{weidemann00} calculates a semi-empirical initial-final mass relation based on 
the available data at the time.  At the low mass end, his relation is constrained 
by an anchor point at $M_{\rm initial}$ = 1~$M_\odot$, $M_{\rm final}$ = 0.55~$M_\odot$ 
which is in good agreement with the mass of the core at the first thermal 
pulse.  The general shape of the relation and possible slope changes are 
discussed in detail.  Interestingly, at the low mass end the data indicate 
that the relation flattens off as the initial-mass scale continues down to 
0.8~$M_\odot$, similar to the core-radius relation shown in 
Figure~\ref{fig:ifmr}.  The NGC~6819 data point at $M_{\rm initial}$ = 1.61~$M_\odot$, 
$M_{\rm final}$ = 0.54~$M_\odot$ is already within a few hundredths of a solar 
mass of the core mass ($\sim$0.5~$M_\odot$ depending on $Z$, see core mass relation 
in Figure~\ref{fig:ifmr}; also Pietrinferni et~al.\ 2004).  The final mass 
of the carbon-oxygen core white dwarf in NGC~6791 is slightly lower than this, 
and equal to the expected mass of white dwarfs in globular clusters 
($M_{\rm final}$ = 0.53~$M_\odot$; Renzini \& Fusi~Pecci 1988; Renzini et~al.\ 
1996; Moehler et~al.\ 2004) with present day turnoffs of $M_{\rm initial}$ = 
0.8~$M_\odot$.  This flattening of the relation suggests an exponential-like behavior at 
low masses.  Unfortunately, such a parameterization would not fit the high 
mass end of the relation very well since those data also appear to show a flattening 
off.  As noted by \cite{weidemann00}, the higher mass stars may in fact form white 
dwarfs that are structurally different in that they have neon/oxygen cores 
instead of carbon-oxygen cores.  Reproducing the relation for stars with 
masses greater than $\sim$4.5~$M_\odot$ can be accomplished with a log function, 
however this would grossly mismatch the masses of the white dwarfs at the low mass 
end.

Lacking a satisfactory functional form of the type discussed above over the entire 
mass range in Figure~\ref{fig:ifmr}, we resort to a simple linear fit as performed in 
the synthesis given by \cite{ferrario05}.  These authors took advantage of 
several recent studies (see earlier references) that have now more than doubled 
the amount of data as compared to the \cite{weidemann00} study (all for 
$M_{\rm initial} >$ 2.75~$M_\odot$).  However, unlike that study, we will use no 
anchor point to fix the relation at the low mass end which is otherwise needed to 
avoid a meaningless slope given the large scatter at intermediate masses 
(Ferrario et~al.\ introduced a point at $M_{\rm initial}$ = 1.1~$M_\odot$, 
$M_{\rm final}$ = 0.55~$M_\odot$).  We also follow the approach introduced by 
\cite{williams06b} and bin the relation so that each star cluster is represented 
as a single point (including Sirius~B).  This has a few advantages.  First, our 
fit will not be over-influenced by the region of the relation with the most data 
points.  Second, the standard deviation in the distribution of masses of white 
dwarfs within a given cluster is a random error on the relation when comparing 
different clusters \citep{williams06b}.  When plotting individual data points, 
an error in the age of a cluster will lead to a systematic offset of all points 
on the relation for that cluster.  Third, our results will be less sensitive to 
any possible peculiar white dwarfs whose initial and final masses are measured 
accurately.  The obvious disadvantage of the binned approach is that a 
given cluster is expected to have white dwarfs with a range of initial and final 
masses and therefore we are throwing away this information.  

The initial-final mass relation based on this binned method is shown in 
Figure~\ref{fig:ifmrcluster}.  We note that the 2$\sigma$ outlier at 
$M_{\rm initial}$ = 5.06~$M_\odot$ is Sirius~B \citep{liebert05b}.  The 
uncertainties in this plot are the standard deviations in the mean initial 
and final mass.  The solid line represents our weighted linear least-squares 
best fit,

\begin{eqnarray*}
M_{\rm final} = (0.109 \pm 0.007)~M_{\rm initial} + 0.394 \pm 0.025~M_\odot. \\
\end{eqnarray*}

\noindent 
Although an ``S'' shaped relation with curvature would provide a better fit 
at both the lower and upper ends, we note that the reduced chi squared per 
degree of freedom is $\chi^2$ = 1.2 in the linear fit and therefore the 
data are well-fit by this simple relation.  If we also include a data point 
at $M_{\rm initial}$ = 0.80 $\pm$ 0.02~$M_\odot$ and $M_{\rm final}$ = 
0.53 $\pm$ 0.02~$M_\odot$ to represent the best current globular cluster 
constraints \citep{renzini96,moehler04}, the relation flattens slightly 
to $M_{\rm final}$ = (0.106 $\pm$ 0.007)~$M_{\rm initial}$ + 
0.409 $\pm$ 0.022~$M_\odot$.  In this case, the $\chi^2$ of the fit is 
1.3.

\section{Discussion and Conclusions}\label{conclusion}

A mapping of the initial mass of a hydrogen burning star to its final remnant 
mass represents an extremely important relation in astrophysics.  Over 
99\% of all stars will end their lives as white dwarfs and expel most of their 
mass into the interstellar medium.  The initial-final mass relation allows us 
to directly integrate this mass loss in a stellar population assuming an 
initial mass function.  Among the many uses of parameterizing this relation is 
a robust estimate for the ages of the Galactic disk and halo.  For example, the 
shape of the white dwarf mass function in the Galactic disk is sharply 
peaked at $\sim$0.6~$M_\odot$ \citep{liebert05b,kepler06}.  The 
initial-final mass relation allows us to reconstruct the distribution of masses 
of the progenitor hydrogen burning stars that formed this peak.  This therefore 
provides an estimate of the age of the Galactic disk, which has now been 
measured to be $\sim$7 -- 9~Gyrs \citep{winget87,wood92,oswalt96,leggett98,hansen02}.  
For a Salpeter initial mass function, and an age of 8~Gyrs for the Galactic disk, 
the shape of the predicted white dwarf mass distribution based on our initial-final 
mass relation is in excellent agreement with the observed mass distribution (i.e., 
the peak location and spread, Kalirai et~al.\ 2008, in preparation).  Similarly, the 
ages of globular clusters in the Galactic halo have been measured to be 
$\sim$12~Gyr by comparing the observed distribution of white dwarfs on 
the cluster CMDs to synthetic cooling sequences produced using an initial 
mass function and an initial-final mass relation \citep{hansen04,hansen07}.  
The relation used most recently in the study of NGC~6397 by Hansen et~al.\ 
(2007) is shown as a dashed line in Figure~\ref{fig:ifmr} and is found to be 
in good agreement with our new low mass constraints (the disagreement at higher 
masses is not important since all of the observed white dwarfs in NGC 6397 
evolved from stars with $M <$ 2~$M_\odot$).

The study of the white dwarf population in NGC~7789, NGC~6819, and NGC~6791 
represents the first time that we have been able to reconstruct this mapping 
for low mass stars such as the Sun, and therefore eliminate the need for an 
indirect anchor point at low masses.  Over half of the total 
number of stars that are produced in a Salpeter-type initial mass function 
now fall within a region of the initial-final mass relation that has some 
constraints.  At high masses, the relation indicates that stars will lose 
80 -- 85\% of their mass through stellar evolution.  However, for stars 
approximately as massive as the Sun, this number drops to $\sim$55\% of 
the initial stellar mass.  

Despite these new data, the importance of the initial-final mass relation 
demands further observations.  To better understand the intrinsic scatter, 
future observations should focus on older clusters with clearly defined white 
dwarf cooling sequences (such as NGC~7789).  By pushing the magnitude limit 
fainter, more massive white dwarfs will be revealed that are likely descendents 
of more massive progenitors.  In this way, multiple initial-final mass relations 
can be constructed over appreciable ranges from the studies of {\it single} 
star clusters whose properties (age, metallicity, binary fraction, etc...) have 
been measured carefully.  Such studies are ideally suited for multi-object 
spectrographs since the white dwarf luminosity function increases as a function 
of magnitude and therefore a large number of objects can be targeted in a single 
exposure.  To truly push the envelope to even lower masses, globular clusters 
should be targeted as well.  The nearest systems, such as M4 and NGC~6397, can 
be studied with 8--10 meter telescopes (e.g., see Moehler et~al.\ 2004).  
In addition to providing constraints 
down to $\sim$0.8~$M_\odot$, the environments of these systems are up to 
100$\times$ more metal-poor than most open clusters and therefore metallicity 
trends can be reliably studied.  At the opposite extreme, rich, young clusters 
can provide unique constraints and push the current high mass limit further.  
This will not only constrain the upper mass limit to white dwarf production, 
but also simultaneously discover the lower mass limit to type II supernova.  
An extrapolation of the present relation to the Chandrasekhar mass suggests 
an initial mass of $\sim$9.5~$M_\odot$, however this is very uncertain given 
the lack of data in this regime.  Finally, there is a paucity of data between 
$M_{\rm initial}$ = 2 -- 2.75~$M_\odot$ which can bridge our new measurements with 
the previous data.  The absence of data points in this region of the relation 
results from a lack of nearby, rich star clusters with an age of $\sim$1~Gyr.  
Fortunately, one such system exists, NGC~2420, and has been shown to possess 
a white dwarf population \citep{vonhippel00} and therefore should be targeted 
in the near future.


\acknowledgements
We gratefully acknowledge P. Bergeron for providing us with his models and 
spectral fitting routines to measure the masses of white dwarfs.  We also 
thank him and James Liebert for taking the time to discuss and help interpret 
the spectra of certain stars.  We wish to thank Robert Eakin for assistance 
with processing of imaging data frames and Don Vandenberg and Leo Girardi for 
their help with interpreting their models.  Finally, we wish to extend our gratitude 
to an anonymous referee for taking the time to prepare a detailed report that has 
resulted in a much improved paper.  JSK is supported by NASA through Hubble 
Fellowship grant HF-01185.01-A, awarded by the Space Telescope Science Institute, 
which is operated by the Association of Universities for Research in Astronomy, 
Incorporated, under NASA contract NAS5-26555. Support for this work was 
also provided by grant HST-GO-10424 from NASA/STScI.  The research of 
HBR is supported by grants from the Natural Sciences and Engineering Research 
Council of Canada. He also thanks the Canada-US Fulbright Program for the award 
of a Fulbright Fellowship.


\end{document}